\documentclass[iop]{emulateapj}
\usepackage{times,natbib,graphicx,amsmath,multirow,xspace}
\usepackage{xcolor}

\shorttitle{X-ray and Radio Observations of 4U~1543$-$624}
\shortauthors{Ludlam et al.}

\newcommand{\nustar}{\textit{NuSTAR}}
\newcommand{\swift}{\textit{Swift}}
\newcommand{\chandra}{\textit{Chandra}}
\newcommand{\nicer}{\textit{NICER}}
\newcommand{\maxi}{\textit{MAXI}}
\newcommand{\integral}{\textit{INTEGRAL}}
\newcommand{\xmm}{{\it XMM-Newton}}
\newcommand{\atca}{ATCA}

\newcommand{\fluxcgs}{ergs~s$^{-1}$~cm$^{-2}$}

\newcommand{\rin}{$R_{\rm in}$}
\newcommand{\rg}{$R_{g}$}

\newcommand{\source}{\mbox{4U~1543$-$624}\xspace}
\newcommand{\xillver}{{\sc xillver$_{\mathrm{CO}}$}}
\newcommand{\lrlx}{${L}_{{r}}$-${L}_{{x}}$}
\newcommand{\lr}{${L}_{{r}}$}

\begin{document}

\title{Observations of the Ultra-compact X-ray Binary \source\ in Outburst with \nicer, \integral, \swift, and \atca}
\author{R.~M.~Ludlam\altaffilmark{1},
L.~Shishkovsky\altaffilmark{2},
P.~M.~Bult\altaffilmark{3},
J.~M.~Miller\altaffilmark{1},
A.~Zoghbi\altaffilmark{1},
T.~E.~Strohmayer\altaffilmark{3},
M.~Reynolds\altaffilmark{1},
L.~Natalucci\altaffilmark{4},
J.~C.~A.~Miller-Jones\altaffilmark{5},
G.~K.~Jaisawal\altaffilmark{6},
S.~Guillot\altaffilmark{7,8},
K.~C.~Gendreau\altaffilmark{3},
J.~A.~Garc\'{i}a\altaffilmark{9,10},
M.~Fiocchi\altaffilmark{4},
A.~C.~Fabian\altaffilmark{11},
D.~Chakrabarty\altaffilmark{12},
E.~M.~Cackett\altaffilmark{13},
A.~Bahramian\altaffilmark{5},
Z.~Arzoumanian\altaffilmark{3},
D. Altamirano\altaffilmark{14}
}
\altaffiltext{1}{Department of Astronomy, University of Michigan, 1085 South University Ave, Ann Arbor, MI 48109-1107, USA}
\altaffiltext{2}{Center for Data Intensive and Time Domain Astronomy, Department of Physics and Astronomy, Michigan State University, East Lansing, MI, USA}
\altaffiltext{3}{X-ray Astrophysics Laboratory, Astrophysics Science Division, NASA/GSFC, Greenbelt, MD 20771, USA}
\altaffiltext{4}{Istituto Nazionale di Astrofisica, INAF-IAPS, via del Fosso del Cavaliere, 100. 00133 Roma, Italy}
\altaffiltext{5}{International Centre for Radio Astronomy Research -- Curtin University, GPO Box U1987, Perth, WA 6845, Australia}
\altaffiltext{6}{National Space Institute, Technical University of Denmark, Elektrovej 327-328, DK-2800 Lyngby, Denmark}
\altaffiltext{7}{CNRS, IRAP, 9 avenue du Colonel Roche, BP 44346, F-31028 Toulouse Cedex 4, France}
\altaffiltext{8}{Universit\'{e} de Toulouse, CNES, UPS-OMP, F-31028 Toulouse, France}
\altaffiltext{9}{Cahill Center for Astronomy and Astrophysics, California Institute of Technology, Pasadena, CA 91125, USA}
\altaffiltext{10}{Remeis Observatory \& ECAP, Universit\"{a}t Erlangen-N\"{u}rnberg, Sternwartstr. 7, D-96049, Bamberg, Germany}
\altaffiltext{11}{Institute of Astronomy, Madingley Road, Cambridge CB3 0HA, UK}
\altaffiltext{12}{MIT Kavli Institute for Astrophysics and Space Research, Massachusetts Institute of Technology, Cambridge, MA 02139, USA}
\altaffiltext{13}{Department of Physics \& Astronomy, Wayne State University, 666 West Hancock Street, Detroit, MI 48201, USA}
\altaffiltext{14}{Department of Physics \& Astronomy, University of Southampton, Highfield, Southampton SO17 1BJ, UK}

\begin{abstract} 
We report on X-ray and radio observations of the ultra-compact X-ray binary \source\ taken in August 2017 during an enhanced accretion episode. 
We obtained \nicer\ monitoring of the source over a $\sim10$ day period during which target-of-opportunity observations were also conducted with \swift, \integral, and \atca. 
Emission lines were measured in the \nicer\ X-ray spectrum at $\sim0.64$ keV and $\sim6.4$ keV that correspond to O and Fe,  respectively. 
By modeling these line components, we are able to track changes in the accretion disk throughout this period.
The innermost accretion flow appears to move inwards from hundreds of gravitational radii (\rg $=GM/c^{2}$) at the beginning of the outburst to $<8.7$ \rg\ at peak intensity. 
We do not detect the source in radio, but are able to place a $3\sigma$ upper limit on the flux density at $27$ $\mu$Jy beam$^{-1}$.
Comparing the radio and X-ray luminosities, we find that the source lies significantly away from the range typical of black holes in the \lrlx\ plane, suggesting a neutron star (NS) primary.
This adds to the evidence that NSs do not follow a single track in the \lrlx\ plane, limiting its use in distinguishing between different classes of NSs based on radio and X-ray observations alone.
\end{abstract}


\section{Introduction}
Ultra-compact X-ray binaries (UCXBs) are low-mass X-ray binaries (LMXBs) with a short orbital period of $<80$ minutes, and typically contain a degenerate stellar companion. 
\source\ is an UCXB with an orbital period of $18.2\pm0.1$ minutes determined from optical photometry \citep{wang04, wang15}. 
The source was first detected by the {\it Uhuru} mission \citep{giacconi72}. 
The absence of hydrogen and helium lines, coupled with emission from carbon and oxygen in the optical spectrum of \source, indicates a degenerate donor star such as a carbon-oxygen white dwarf \citep{nelemans03}. 

The nature of the compact accretor in this system is unknown, but is often assumed to be a neutron star (NS). 
The \textit{Monitor of All-sky X-ray Image} (\maxi) recently detected a Type-I X-ray burst \citep{serino18} consistent with the position of \source\footnote[1]{http://maxi.riken.jp/alert/novae/8151027492/8151027492.htm}, although the instrument's spatial resolution of 1.5$^{\circ}$ \citep{matsuoka09} prevents a definitive conclusion regarding the event's origin since there are other X-ray sources nearby. If the burst indeed arose from \source, then this would irrefutably confirm the system contains a NS. Additionally, this would place \source\ within a subset of unique UCXBs that show helium-powered X-ray bursts even though they have degenerate C/O donors and their optical spectra lack evidence of He (4U~0614+091: \citealt{kuulkers10}, IGR J17062$-$6143: \citealt{strohmayer18}).

\begin{figure*}[t]
\begin{center}
\includegraphics[width=\textwidth]{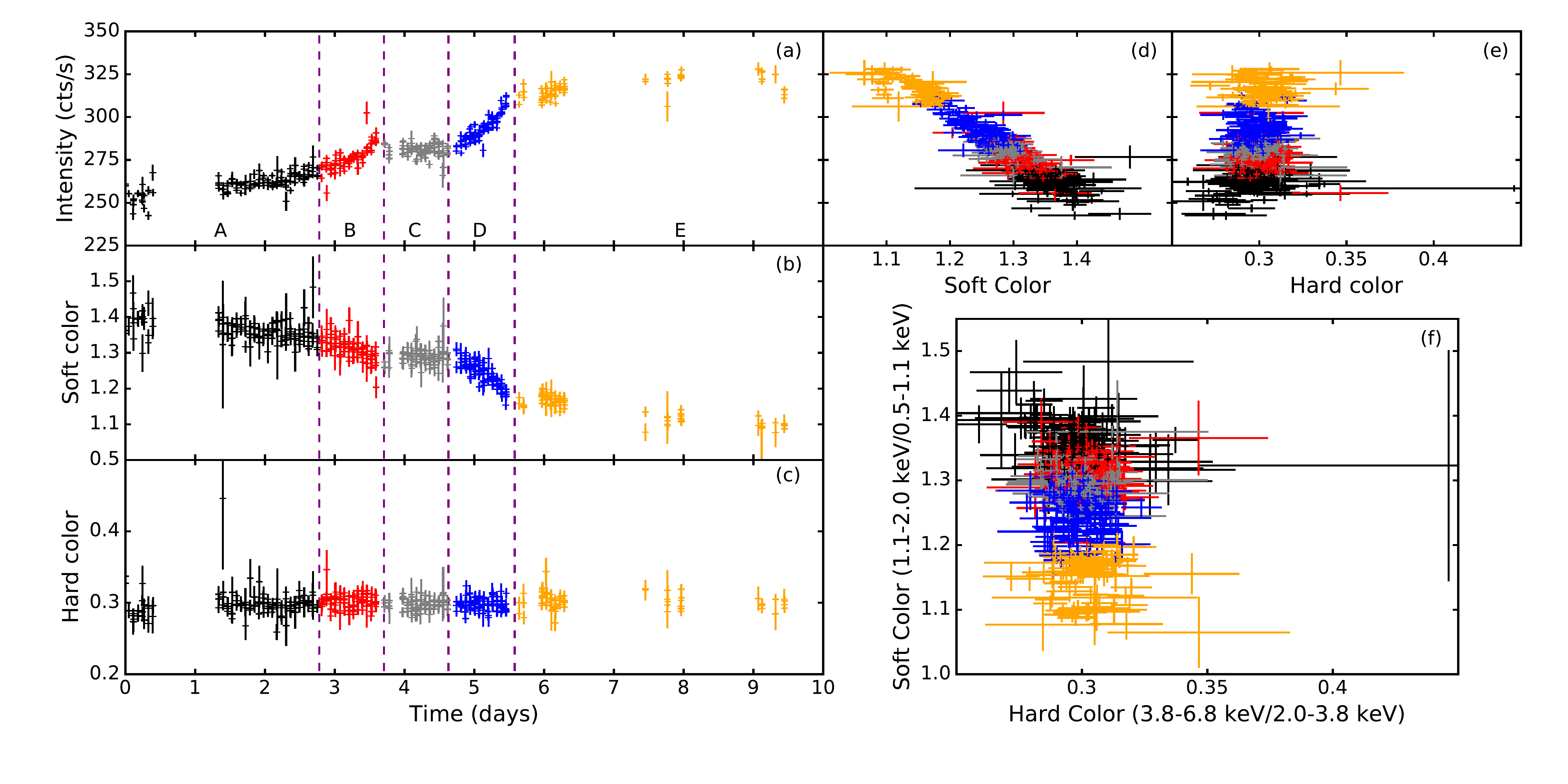}
\caption{ Panel (a): The $0.5-6.8$ keV \nicer\ light-curve of \source\ starting on 2017 Aug 16. The vertical dashed purple lines indicate the time intervals in which spectra were extracted.  Data were binned to 128 s. Panel (b): The soft color \mbox{($1.1-2.0$ keV / $0.5-1.1$ keV)} evolution of the source.  Panel (c): The hard color \mbox{($3.8-6.8$ keV / $2.0-3.8$ keV)}  evolution throughout the outburst.   Panels (d) and (e): Soft color and hard color intensity diagrams. Panel (f): Soft color versus hard color diagram. The single outlier point in interval A is an artifact of the 128 s time binning. }
\label{fig:intervals}
\end{center}
\end{figure*} 

Further support that \source\ contains a NS  can be obtained from observing the source in both radio and X-rays. 
For LMXBs, the general view is that the X-ray emission tracks the accretion inflow closest to the compact object, while radio emission represents an outflow in the form of a relativistic jet \citep{fender03}. 
The exact mechanism creating these jets is not confirmed and may vary with different systemic properties, but is largely accepted to involve the acceleration of material by the magnetic field of the compact object or inner accretion disk, creating the synchrotron radio emission that has been observed in many LMXBs (e.g., \citealt{blandford77}; \citealt{blandford82}; \citealt{tudor17}).
Observing the radio and X-ray luminosity simultaneously can be advantageous in discerning between a NS or BH accretor, since NSs are typically  $\sim30$ times fainter in the radio than BHs at the same X-ray luminosity \citep{migliari06}.
However, some caution should be maintained in utilizing this diagnostic tool given that certain classes of NSs, namely transitional and accreting millisecond X-ray pulsars (tMXPs and AMXPS), can be nearly as radio-luminous as the radio-faint black holes \citep{deller15}.

In many LMXBs, hard X-ray photons incident on the surrounding material in the accretion disk will result in emission features from the photons being reprocessed. 
These emission features are broadened by strong Doppler, special relativistic, and general relativistic effects from proximity to the compact accretor \citep{fabian00}. 
The strength of these effects can therefore be used to determine how close the accretion disk is to the NS or BH.

\source\ was first reported to exhibit a broad O {\sc viii} Ly$\alpha$ emission feature at $\sim0.7$ keV by \citet{madej11} when observed with the high-resolution spectrographs onboard \xmm\ and \chandra. 
Previous evidence for an emission feature near 0.7 keV existed in {\it ASCA} and {\it BeppoSAX} observations (\citealt{juett01}; \citealt{schultz03}), but these early spectra could also be described without invoking O emission if there was an overabundance in neon that increased absorption along the line of sight \citep{juett01}. This generated uncertainty surrounding the identification of this feature.
A similar O feature was reported with \chandra\ and \xmm\ by \citet{juett03}, but early calibration uncertainties hindered a definitive claim.

In addition to O {\sc viii} Ly$\alpha$, there was also evidence of a possible Fe K$\alpha$ feature at $\sim6.6$ keV concurrent with the detection of O {\sc viii} Ly$\alpha$ \citep{madej11}. 
This was not the first time that the Fe line had been detected in this system, since it was present in prior {\it RXTE} and {\it EXOSAT} data \citep{schultz03}, but \citet{madej11} were the first to claim that both features were present in the X-ray spectrum. 
The O {\sc viii} feature was found to be more intense relative to the continuum than Fe~K. 
For a typical accretion disk of solar abundance, O {\sc viii} is the second strongest feature after Fe~K. 
In order for O to be the most prominent feature, the disk would need a significant overabundance of oxygen \citep{ballantyne02}, further supporting that the donor star in the system is a CO or ONe white dwarf \citep{madej11}.

\citet{madej14} presented an X-ray spectral analysis using a preliminary version of a new reflection model tailored to accommodate the typical elemental abundances in UCXBs known as \xillver. 
This model has negligible H and He abundances, variable abundance of C and O, and 10 times solar abundance for all other elemental species, but had only a limited number of grid points (i.e., large steps between parameter values) at the time.
Spectral modeling using this initial \xillver\ grid on \source\ indicated an inclination of $i\sim65^{\circ}$ and an inner disk radius $<7.4$ \rg\ (where $R_{g}=GM/c^{2}$).

In early August 2017, the \swift/BAT detected increased activity from \source\ in the $15-50$ keV energy band \citep{ludlam17}. 
We requested \nicer\ monitoring of the source in order to track any changes in the system as the X-ray intensity increased. 
Additionally, we secured target-of-opportunity observations with \swift, \integral, and \atca.
We present the results of our X-ray observations throughout this period of enhanced accretion and reanalyze the radio data that was initially reported in \citet{ludlam17} below. 

\begin{table*}[t!]
\caption{Spectral Modeling of X-ray Observations}
\label{tab:spec} %
\begin{center}
\begin{tabular}{lcccccc}
\hline


Model 1 & Parameter & Interval A & Interval B & Interval C & Interval D & Interval E \\
\hline
{\sc tbabs} 

& $\mathrm{N}_{\mathrm{H}}$ ($10^{21}$ cm$^{-2}$) 
&  \multicolumn{5}{c}{$2.92_{-0.03}^{+0.02}$}\\

{\sc constant} 

& C 
& $0.93\pm0.03$ 
& ... 
& ... 
& ... 
& $0.33\pm0.03$ \\

{\sc diskbb}

& kT ($10^{-2}$ keV) 
& $4.9_{-0.4}^{+0.3}$ 
& $8.7_{-0.8}^{+2.8}$ 
& $9.3_{-0.6}^{+0.4}$ 
& $12.2_{-0.3}^{+0.4}$ 
& $16.1\pm0.3$ \\

& norm$_{\mathrm{disk}}$ ($10^{5}$) 
& $150_{-70}^{+222}$ 
& $1.9\pm1.5$ 
& $1.2_{-0.3}^{+0.6}$ 
& $0.36\pm0.06$
& $0.20\pm0.02$ \\

{\sc bbody} 
& kT (keV) 
& $0.66\pm0.01$ 
& $0.67_{-0.04}^{+0.02}$ 
& $0.73\pm0.01$ 
& $0.74_{-0.02}^{+0.01}$ 
& $0.65_{-0.01}^{+0.02}$\\

& norm$_{\mathrm{bb}}$ ($10^{-3}$) 
& $1.04_{-0.03}^{+0.04}$ 
& $0.89_{-0.03}^{+0.05}$ 
& $1.06\pm0.04$ 
& $1.17\pm0.04$ 
& $1.22\pm0.06$ \\

{\sc powerlaw} 
& $\Gamma$ 
& $1.74\pm0.01$ 
& $1.73_{-0.02}^{+0.01}$ 
& $1.77\pm0.01$ 
& $1.79\pm0.01$ 
& $1.71\pm0.01$ \\

& norm$_{\mathrm{pl}}$ ($10^{-1}$) 
& $1.46\pm0.01$ 
& $1.58_{-0.03}^{+0.01}$ 
& $1.61\pm0.01$ 
& $1.66_{-0.02}^{+0.01}$ 
& $1.59\pm0.03$ \\


&$\chi^{2}$ (dof) & \multicolumn{5}{c}{8306.01 (4668)}\\ 
\hline


Model 2 & Parameter & Interval A & Interval B & Interval C & Interval D & Interval E\\
\hline
{\sc tbabs} 

& $\mathrm{N}_{\mathrm{H}}$ ($10^{21}$ cm$^{-2}$) 
& \multicolumn{5}{c}{$3.06_{-0.02}^{+0.03}$}\\

{\sc constant} 

& C 
& $0.93\pm0.03$ 
& ... 
& ... 
& ... 
& $0.44_{-0.04}^{+0.05}$ \\

{\sc diskbb} 

& kT ($10^{-2}$ keV) 
& $4.6\pm0.2$  
& $5.7_{-0.2}^{+0.3}$ 
& $5.6\pm0.2$ 
& $5.7\pm0.2$ 
& $10.6_{-0.6}^{+0.2}$\\

& norm$_{\mathrm{disk}}$ ($10^{5}$) 
& $456_{-150}^{+278}$ 
& $68_{-24}^{+25}$ 
& $88_{-19}^{+36}$ 
& $81_{-17}^{+24}$ 
& $1.4_{-0.1}^{+0.5}$\\

{\sc bbody} 

& kT (keV) 
& $0.69\pm0.01$ 
& $0.74\pm0.02$ 
& $0.80_{-0.01}^{+0.02}$ 
& $0.83_{-0.01}^{+0.02}$ 
& $0.81_{-0.02}^{+0.04}$ \\

& norm$_{\mathrm{bb}}$ ($10^{-3}$)
& $1.07_{-0.05}^{+0.03}$ 
& $0.94_{-0.09}^{+0.06}$ 
& $1.24_{-0.08}^{+0.06}$ 
& $1.36_{-0.07}^{+0.09}$ 
& $0.98_{-0.09}^{+0.08}$\\

{\sc powerlaw} 

& $\Gamma$
& $1.77\pm0.01$ 
& $1.78\pm0.01$ 
& $1.83\pm0.01$ 
& $1.86\pm0.01$ 
& $1.82\pm0.01$\\

& norm$_{\mathrm{pl}}$ ($10^{-1}$) 
& $1.51_{-0.01}^{+0.02}$ 
& $1.78_{-0.01}^{+0.02}$ 
& $1.66\pm0.01$ 
& $1.74\pm0.01$ 
& $1.86_{-0.01}^{+0.02}$ \\

{\sc diskline}$_{1}$ 

& E$_{\mathrm{O}}$ (keV) 
& $0.637_{-0.009}^{+0.004}$ 
& $0.637_{-0.008}^{+0.010}$ 
& $0.646_{-0.010}^{+0.008}$ 
& $0.632_{-0.008}^{+0.006}$ 
& $0.659_{-0.008}^{+0.005}$ \\

& $q$ & \multicolumn{5}{c}{$-2.36_{-0.07}^{+0.05}$} \\

& $R_{\mathrm{in}}$ (\rg)
& $942_{-754}^{+30}$ 
& $46_{-19}^{+65}$ 
& $44_{-19}^{+39}$ 
& $7.9_{-1.3}^{+2.5}$ 
& $7.3_{-1.0}^{+1.4}$ \\

& $i$ ($^{\circ}$)
& \multicolumn{5}{c}{$74_{-4}^{+6}$} \\

& norm$_{\mathrm{line}}$ ($10^{-2}$) 
& $0.19_{-0.02}^{+0.04}$ 
& $0.40_{-0.07}^{+0.08}$ 
& $0.45_{-0.06}^{+0.10}$ 
& $1.2\pm0.1$ 
& $1.5_{-0.1}^{+0.2}$ \\

& EW$_{\mathrm{O}}$ (eV)
& $5.0_{-0.5}^{+1.1}$
& $10_{-1.75}^{+2.0}$
& $11_{-1.5}^{+2.4}$
& $27\pm2.3$
& $33_{-2.2}^{+4.4}$\\

{\sc diskline}$_{2}$ 

& E$_{\mathrm{Fe}}$ (keV) 
& $6.42_{-0.01}^{+0.28}$ 
& $6.7\pm0.2$ 
& $6.41_{-0.01}^{+0.13}$ 
& $6.41_{-0.01}^{+0.06}$ 
& $6.48_{-0.08}^{+0.27}$ \\

& $R_{\mathrm{in}}$ (\rg)
& $28_{-2}^{+26}$ 
& $17\pm11$ 
& $9_{-2}^{+6}$ 
& $12_{-5}^{+7}$ 
& $10.2_{-3.4}^{+12.5}$ \\

& norm$_{\mathrm{line}}$ ($10^{-3}$) 
& $0.6_{-0.2}^{+0.1}$ 
& $0.9_{-0.5}^{+0.3}$ 
& $1.2\pm0.3$ 
& $1.3\pm0.3$ 
& $0.8\pm0.3$ \\

& EW$_{\mathrm{Fe}}$ (eV)
& $105_{-35}^{+18}$
& $160_{-89}^{+53}$
& $210\pm53$
& $228\pm53$
& $123\pm46$ \\

&$\chi^{2}$ (dof) & \multicolumn{5}{c}{5903.06 (4636)}\\ 


\hline
Model 3 & Parameter & Interval A & Interval B & Interval C & Interval D & Interval E\\
\hline
{\sc tbabs} 

& $\mathrm{N}_{\mathrm{H}}$ ($10^{21}$ cm$^{-2}$) 
& \multicolumn{5}{c}{$3.05_{-0.02}^{+0.04}$}\\

{\sc constant} 

& C 
& $0.93\pm0.03$ 
& ... 
& ... 
& ... 
& $0.45_{-0.04}^{+0.05}$ \\

{\sc diskbb} 

& kT ($10^{-2}$ keV) & $4.7_{-0.2}^{+0.1}$  & $5.7_{-0.4}^{+0.6}$ & $5.5_{-0.3}^{+0.1}$ & $5.6\pm0.2$ & $10.1_{-0.4}^{+0.2}$\\

& norm$_{\mathrm{disk}}$ ($10^{5}$) & $400_{-90}^{+200}$ & $74_{-43}^{+74}$ & $102_{-22}^{+77}$ & $97_{-26}^{+47}$ & $1.74_{-0.2}^{+0.5}$\\

{\sc bbody} 

& kT (keV) & $0.68_{-0.01}^{+0.02}$ & $0.72\pm0.02$ & $0.79_{-0.01}^{+0.02}$ & $0.83_{-0.01}^{+2}$ & $0.81_{-0.01}^{+0.03}$ \\

& norm$_{\mathrm{bb}}$ ($10^{-3}$) & $1.00_{-0.03}^{+0.04}$ & $0.88_{-0.06}^{+0.05}$ & $1.17_{-0.04}^{+0.07}$ & $1.40_{-0.06}^{+0.8}$ & $1.00_{-0.07}^{+0.08}$\\

{\sc powerlaw} 

& $\Gamma$ & $1.76\pm0.01$ & $1.77\pm0.01$ & $1.82\pm0.01$ & $1.87\pm0.01$ & $1.83\pm0.01$\\

& norm$_{\mathrm{pl}}$ ($10^{-1}$) & $1.51_{-0.01}^{+0.02}$ & $1.64_{-0.01}^{+0.02}$ & $1.66_{-0.01}^{+0.02}$ & $1.74_{-0.01}^{+0.02}$ & $1.87_{-0.01}^{+0.02}$ \\

{\sc diskline}$_{1}$ 

& E$_{\mathrm{O}}$ (keV) & $0.637_{-0.017}^{+0.003}$ & $0.63_{-0.02}^{+0.01}$ & $0.63\pm0.01$ 
& $0.63\pm0.01$ 
& $0.66\pm0.01$ \\

& $q$ & \multicolumn{5}{c}{$-2.34_{-0.07}^{+0.05}$} \\

& $R_{\mathrm{in}}$ (\rg) & $515\pm455$ & $32_{-16}^{+22}$ & $18_{-5}^{+11}$ & $6.94_{-9.4}^{+3.0}$ & $6.1_{-0.1}^{+2.1}$ \\

& $i$ ($^{\circ}$)
& \multicolumn{5}{c}{$70_{-1}^{+10}$} \\

& norm$_{\mathrm{line}}$ ($10^{-2}$) 
& $0.20_{-0.02}^{+0.07}$ 
& $0.4\pm0.1$ 
& $0.55_{-0.06}^{+0.09}$  
& $1.2\pm0.1$ 
& $1.6\pm0.1$ \\

& EW$_{\mathrm{O}}$ (eV)
& $5.5_{-0.5}^{+1.9}$
& $11\pm3$
& $13.3_{-1.4}^{+2.2}$
& $27\pm2$
& $35\pm2$\\

{\sc diskline}$_{2}$ 

& E$_{\mathrm{Fe}}$ (keV)  & $6.41_{-0.01}^{+0.28}$ & $6.7\pm0.2$ & $6.41_{-0.01}^{+0.20}$ 
& $6.41_{-0.01}^{+0.05}$ 
& $6.41_{-0.01}^{+0.21}$ \\

& norm$_{\mathrm{line}}$ ($10^{-3}$)  
& $0.2_{-0.1}^{+0.3}$  
& $0.4_{-0.1}^{+0.2}$ 
& $0.8\pm0.2$ 
& $1.5\pm0.2$ 
& $0.9\pm0.3$ \\

& EW$_{\mathrm{Fe}}$ (eV)
& $26_{-13}^{+39}$
& $75_{-19}^{+37}$
& $142\pm36$
& $267\pm36$
& $143\pm48$\\

&$\chi^{2}$ (dof) & \multicolumn{5}{c}{5936.04 (4641)}\\ 
\hline
\end{tabular}
\end{center}

\medskip
Note.---  Errors are reported at the 90\% confidence level and calculated from Markov Chain Monte Carlo (MCMC) of chain length $10^{6}$. \nicer\ is fit in the $0.4-9.5$ keV energy band,  \integral\ is fit in the $26-100$ keV band, and \swift\ is considered in the $0.3-8.5$ keV band.  The $\chi^{2}$ (dof) reported refers to the simultaneous fitting of all spectra. The multiplicative constant is used on the \swift\ data in Interval A and \integral\ data in Interval E  with all other parameters tied to the \nicer\ data in that respective interval. The outer disk radius is fixed at 1000 \rg. Inclination ($i$) and emissivity index ($q$) are tied between the two disk line components. \rin\ is tied between each {\sc diskline} component in Model 3.
\end{table*}

\section{Observations and Data Reduction}

\subsection{\nicer}
The {\it Neutron Star Interior Composition Explorer} (\nicer; \citealt{gendreau12}) onboard the International Space Station is comprised of 56 ``concentrator" optics and silicon drift detector pairs that each collect X-rays in the $0.2-12$ keV range.
The 52 operational detectors provide a collecting area of $\sim1900$ cm$^{2}$ at 1.5 keV. \nicer\ observed \source\ over a $\sim$10 day period in 2017 August (ObsIDs 1050060101-1050060113) for a cumulative exposure time of $\sim86.2$ ks. 
Data were reduced using {\sc nicerdas} 2018-10-07\_V005. 
We created good time intervals (GTIs) using {\sc nimaketime} to select events that occurred when the space weather index KP~$<$~5 to avoid periods of time with high particle background due to geomagnetic storms and magnetic cut-off rigidity of COR\_SAX~$>$~4 to ensure that high particle radiation intervals coincident with the Earth's auroral zones are removed.
These GTIs were applied to the data using {\sc niextract-events} selecting PI energy channels between 25 and 1200, inclusive, and EVENT\_FLAGS=bxxx1x000.

The event files for each observation were loaded into {\sc xselect} and combined to create light-curves and spectra.
Light-curves were extracted in 5 different energy bands (super soft: $0.5-1.1$~keV, soft: $1.1-2.0$~keV, intermediate: $2.0-3.8$~keV, hard: $3.8-6.8$~keV, and full: $0.5-6.8$~keV) and binned to 128 s for color analysis as per \citet{bult18}. 
Figure \ref{fig:intervals} shows the \nicer\ light-curve, soft and hard color evolution, color-intensity, and color-color diagrams. 
We extract five spectra based on changes in the light-curve and color diagram, which are labeled A--E. 
These will be referred to as intervals A--E in later sections. 
The spectra are normalized to instrumental residuals created from the Crab Nebula (see \citealt{ludlam18} for more details).
Backgrounds were generated based upon the same filtering criteria using {\it RXTE} ``blank sky" field 5 \citep{jahoda06}. 
The cleaned source spectra have exposure times of 11.4~ks, 8.45~ks, 10.3~ks, 10.2~ks, 7.96~ks for intervals A-E, respectively, after filtering. 
We use the publicly available ARF and RMF instrument response files from 2017 June 1 for these observations. 

\subsection{\swift}
4U 1543-624 has been observed on numerous occasions by the Neil Gehrels Swift Observatory (\swift). 
There was a single XRT observation (ObsID: 00010238010) in Windowed Timing mode on 2017 Aug 18 that overlapped with the \nicer\ data during interval~A.
The raw \textit{Swift} observation files were downloaded from HEASARC and reprocessed using \textsc{xrtpipeline} and  CALDB version 20180710. 
Although the observation is short ($\sim200$ s), the source is detected with a net count rate of $18.05\pm0.32$ ct/s, which is well below the threshold for photon pile-up in Windowed Timing mode.

\subsection{\integral}
The \integral\ observations occurred during interval E of the \nicer\ monitoring for 23.9 ks. 
The \integral\//IBIS \citep{ubertini03} data for this observation were processed using the standard off-line Scientific Analysis (OSA v10.2) software released by the \integral\/ Scientific Data Centre \citep{courvoisier03} in order to obtain a spectrum. 
These runs were performed with the AVES cluster, designed to optimize performance and disk storage needed for the \integral\ data analysis by \citet{federici10}.

\subsection{\atca}
4U 1543-624 was observed by the Australia Telescope Compact Array (ATCA) on 2017 Aug 23, from 11:24:59.9 to 16:09:19.9 UTC, with  $\approx$ 3.25 hrs on source (project code CX392). 
This falls within interval E of the \nicer\ monitoring. 
The 4cm CABB receiver was used, which was set up with two frequency subbands of 2048 MHz bandwidth each, centered at 5.5 and 9.0 GHz.
The ATCA array was in the 1.5A configuration, which has a maximum baseline length of 4.47 km. 
The flux and bandpass calibrator used for this observation was 1934-638, while 1554-64 was used as a phase calibrator.

\begin{figure} 
\begin{center}
\includegraphics[width=0.48\textwidth]{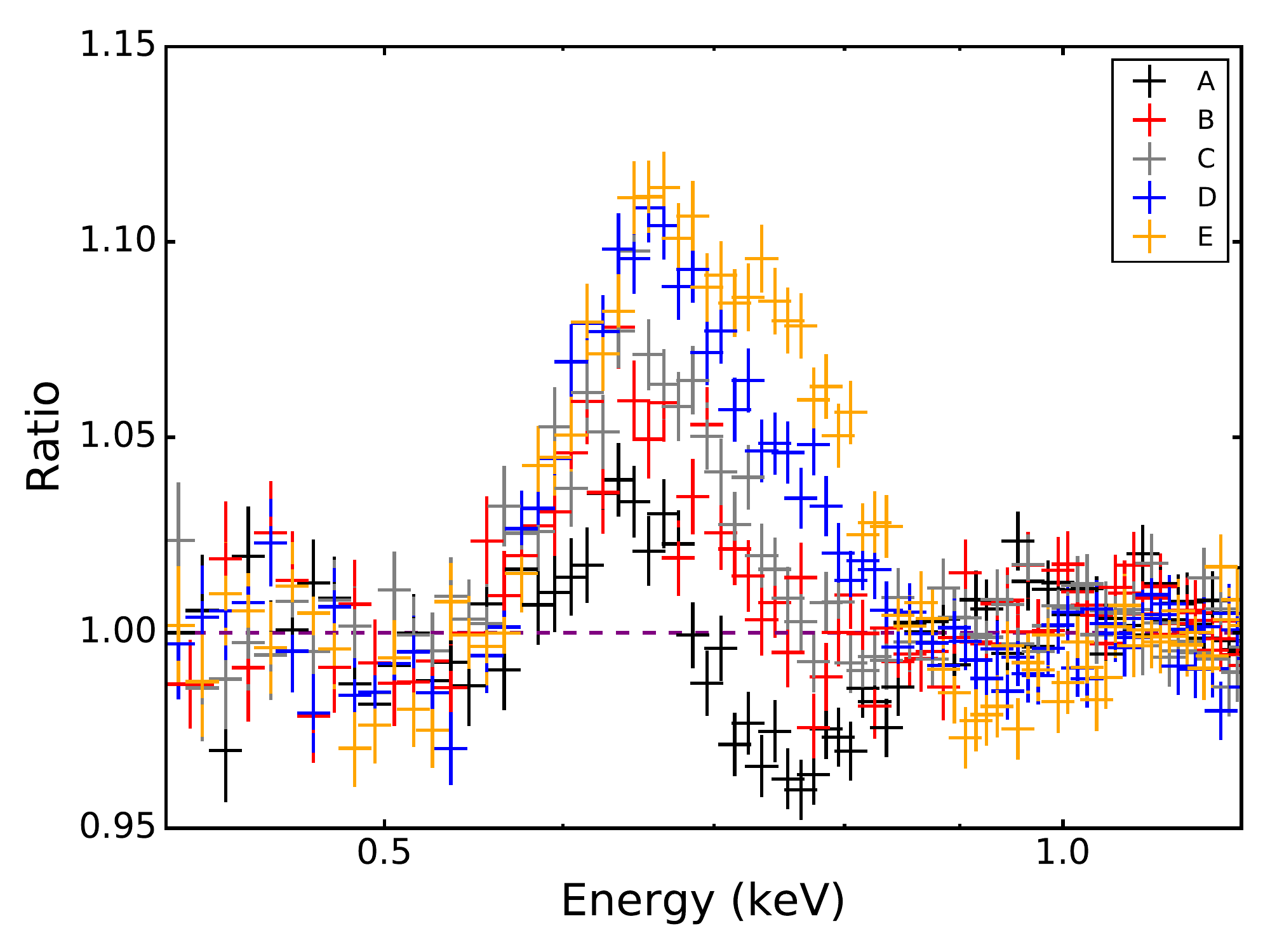}
\includegraphics[width=0.49\textwidth]{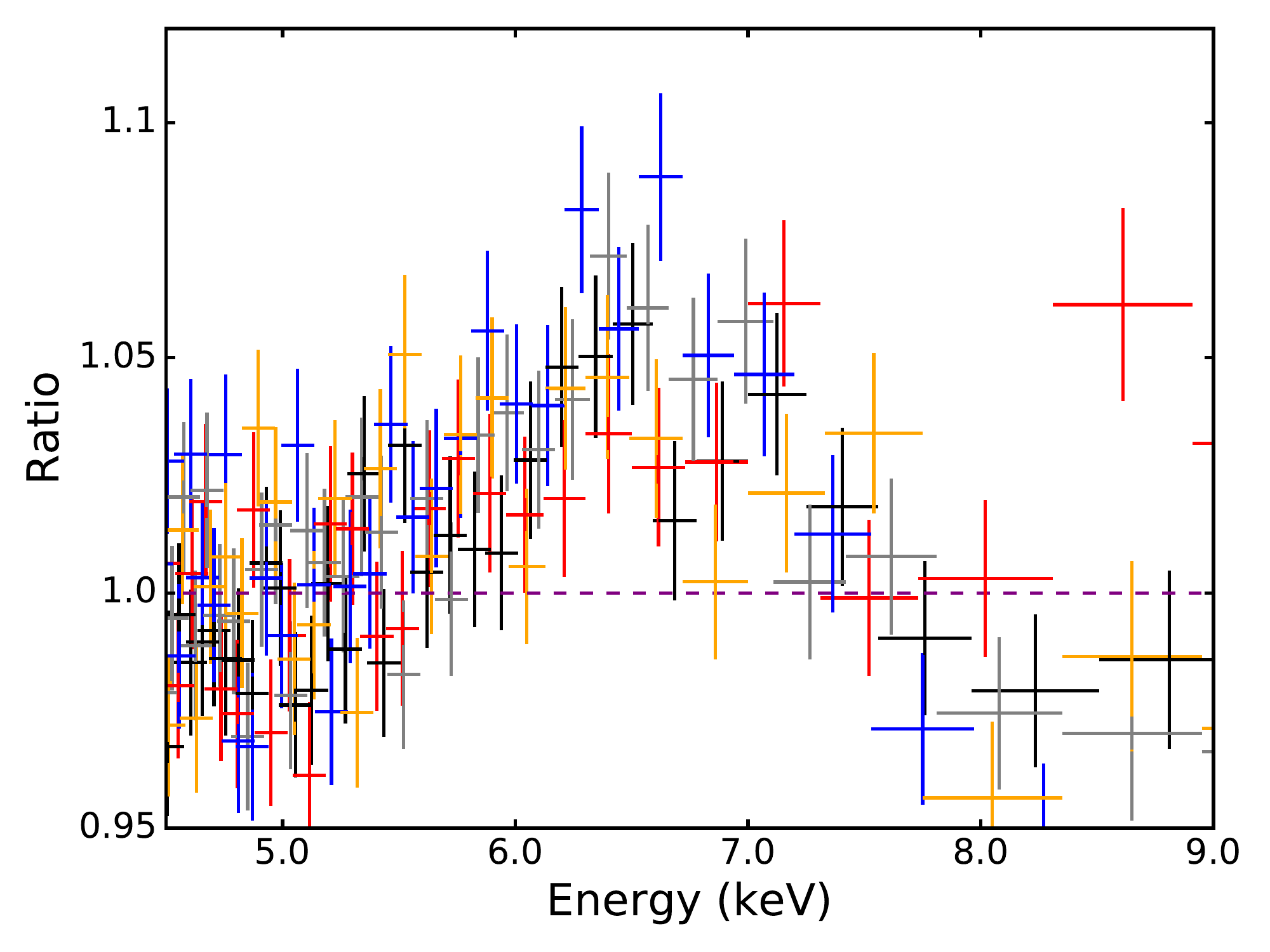}
\caption{Ratio of \nicer\ data to continuum model in each interval for the O band (top) and Fe K band (bottom). The continuum is modeled by an absorbed multi-temperature accretion disk, single-temperature blackbody, and power-law. The apparent absorption feature between $0.7-0.8$ keV in interval A is likely due to an O {\sc vii} edge that disappears as the ionization state of the material changes (see text for more detail).}
\label{fig:lines}
\end{center}
\end{figure}

Data for each frequency band were separately reduced according to standard procedures with Miriad \citep{sault95}. 
The target visibilities were then imported into the Common Astronomy Software Application (CASA; \citealt{mcmullin07}) to be imaged. 
Each 5.5 GHz and 9.0 GHz baseband was imaged separately, with an additional image created combining both datasets in the uv-plane having a central frequency of 7.25 GHz. 
We used the CASA task {\tt tclean}, selecting Briggs weighting with a robust parameter of 1, as well as  nterms=2 to account for non-zero spectral indices in other field sources. 

The synthesized beams at 5.5 GHz, 7.25 GHz, and 9.0~GHz were 6.3\arcsec $\times$ 2.1\arcsec, 5.4\arcsec $\times$ 1.8\arcsec, and 4.1\arcsec $\times$ 1.4\arcsec, respectively.
The local sensitivities achieved were: 13$\ \mu$Jy beam$^{-1}$ at 5.5~GHz; 9$\ \mu$Jy beam$^{-1}$ at 7.25~GHz; 10$\ \mu$Jy beam$^{-1}$ at 9.0~GHz. 
These are slightly higher than the expected theoretical sensitivities, due to several bright field sources in the primary beam side-lobes. 
In any case, \source is clearly not detected in any of our images.
We derive a 3$\sigma$ upper limit on the radio luminosity of \source using the corresponding 3$\sigma$ sensitivity of our deepest image (7.25 GHz): 27$\ \mu$Jy beam$^{-1}$. 

\section{X-ray Analysis and Results}
\subsection{Spectral}
We use {\sc xspec} \citep{arnaud96} version 12.10.1 for our spectral analysis. 
Parameter uncertainties are reported at the 90\% confidence level. 
These are determined from Monte Carlo Markov Chains of length $10^{6}$ with an equal burn-in length using the `chain' command. 
\nicer\ spectra are modeled in the $0.4-9.5$~keV band since the effective collecting area drops sharply outside of this region \citep{ludlam18}. 
The \swift/XRT data are considered in the $0.3-8.5$~keV band since the spectrum becomes quickly background-dominated above this energy. 
The \integral/IBIS observation is fit in the $26-100$~keV band. 
In order to account for differences in calibration between missions, a multiplicative constant is allowed to float between the \swift\ and \nicer\ data in interval~A and \integral\ and \nicer\ during interval~E. 

\begin{figure} 
\begin{center}
\includegraphics[width=0.48\textwidth]{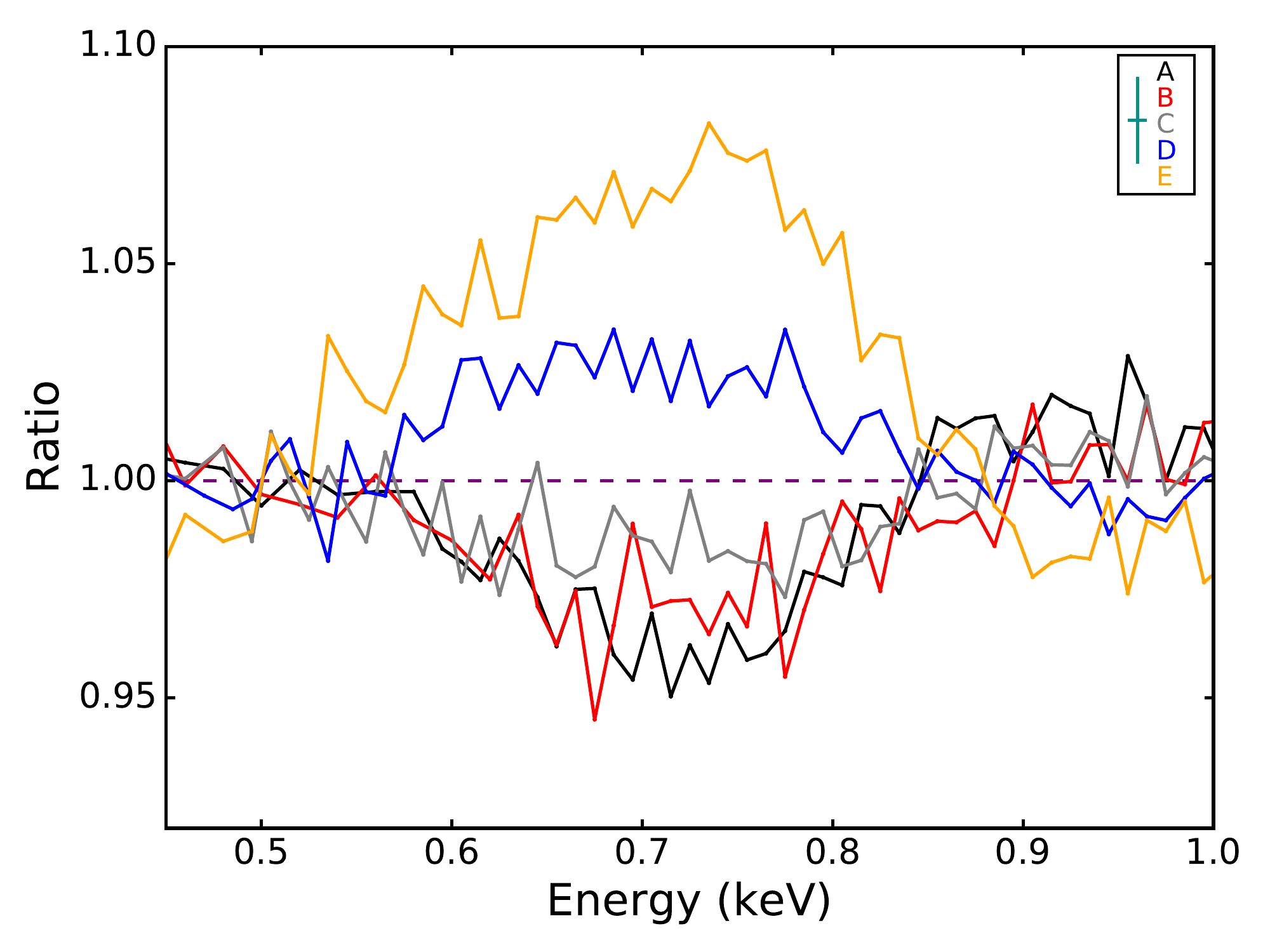}
\caption{Ratio of the \nicer\ spectrum in each interval normalized to the time average spectrum. This demonstrates a clear change in the O line profile that is independent of the calibration correction. The teal point indicates the errors in the x and y directions. Values larger than unity indicate that emission is stronger in that region, whereas values less than unity indicate that the emission is weaker than the mean O profile.}
\label{fig:Otan}
\end{center}
\end{figure}

\begin{figure} 
\begin{center}
\includegraphics[width=0.34\textwidth, angle=270]{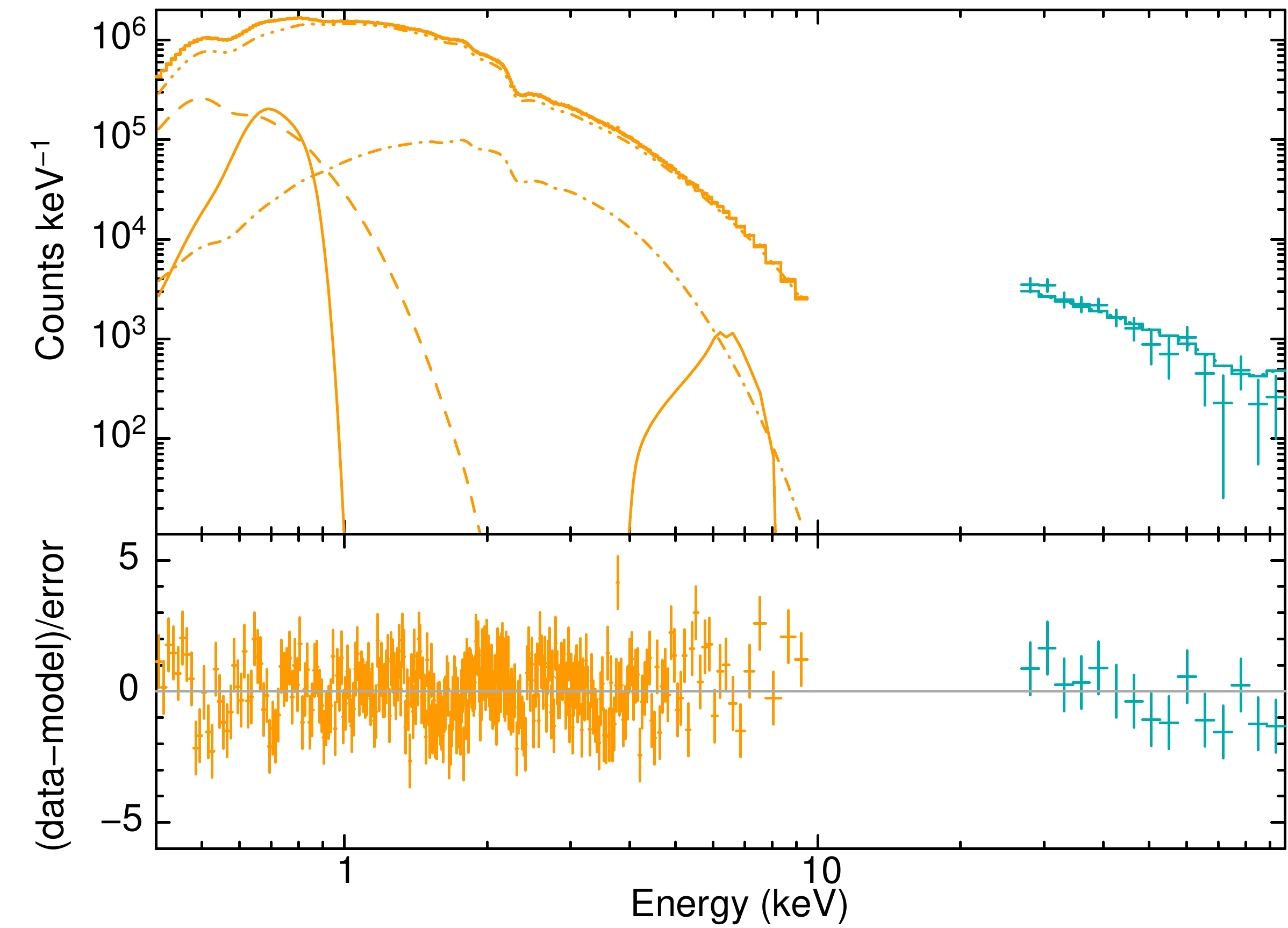}
\caption{Spectral modeling of \nicer\ (orange) and \integral\ (teal) data in interval E with the residuals divided by the errors shown in the lower panel. The dashed line indicates the multi-temperature blackbody component. The dot-dashed line corresponds to the single-temperature blackbody. The dot-dot-dot-dashed line indicates the power-law component. The solid lines indicate the {\sc diskline} components for O and Fe emission. This is representative of the broadband modeling applied to each interval, but only one interval is shown for simplicity. Data were rebinned for clarity.}
\label{fig:spectra}
\end{center}
\end{figure}

\begin{figure} 
\begin{center}
\includegraphics[width=0.48\textwidth]{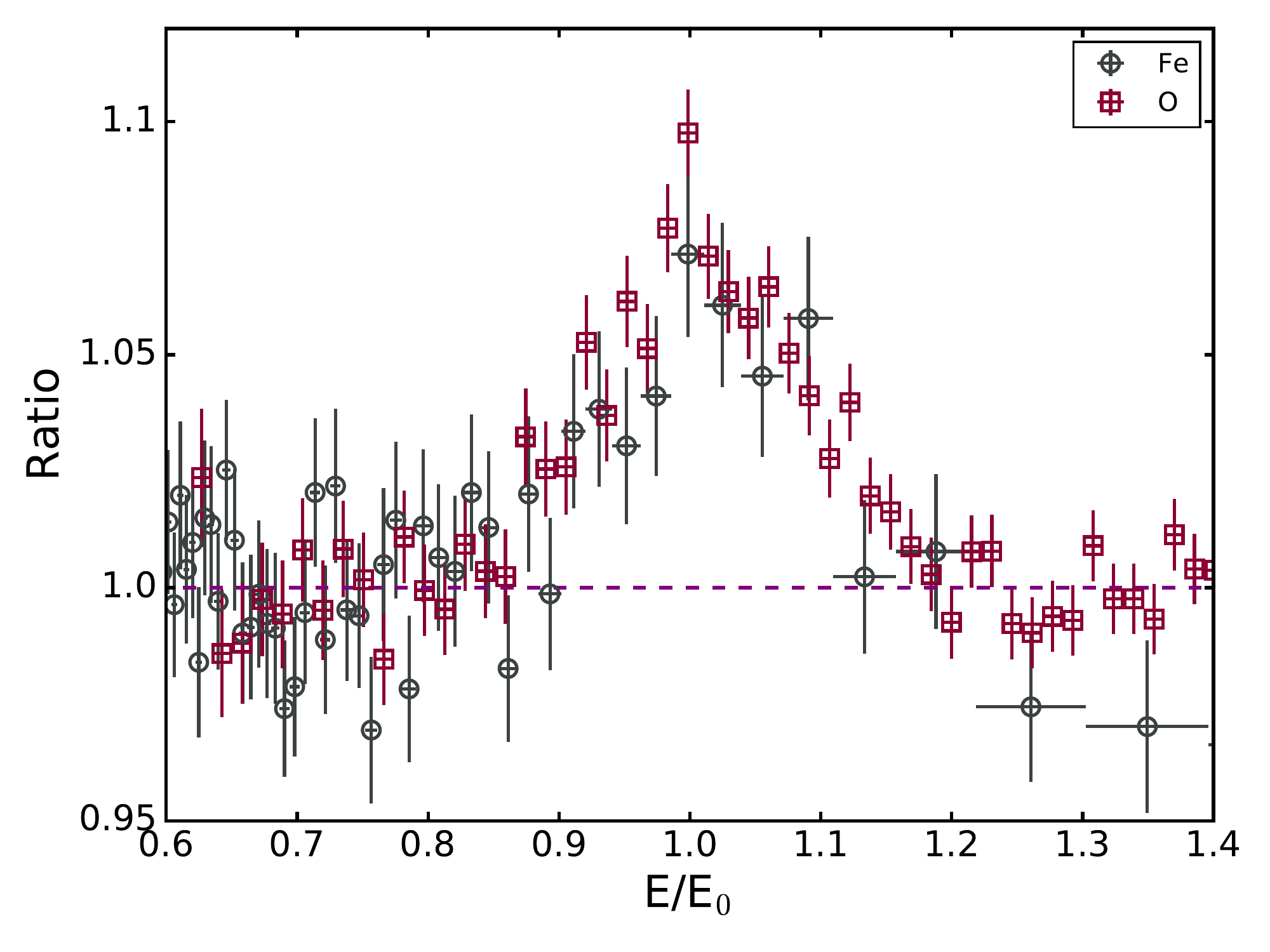}
\caption{The Fe and O line profiles in \nicer\ from interval C plotted in velocity space,  referred to the line centroid energies reported in Table \ref{tab:spec}. The lines show similar broadening profiles indicating a common origin within the accretion disk.}
\label{fig:EE0}
\end{center}
\end{figure}

\begin{figure*}
\begin{center}
\includegraphics[width=0.48\textwidth]{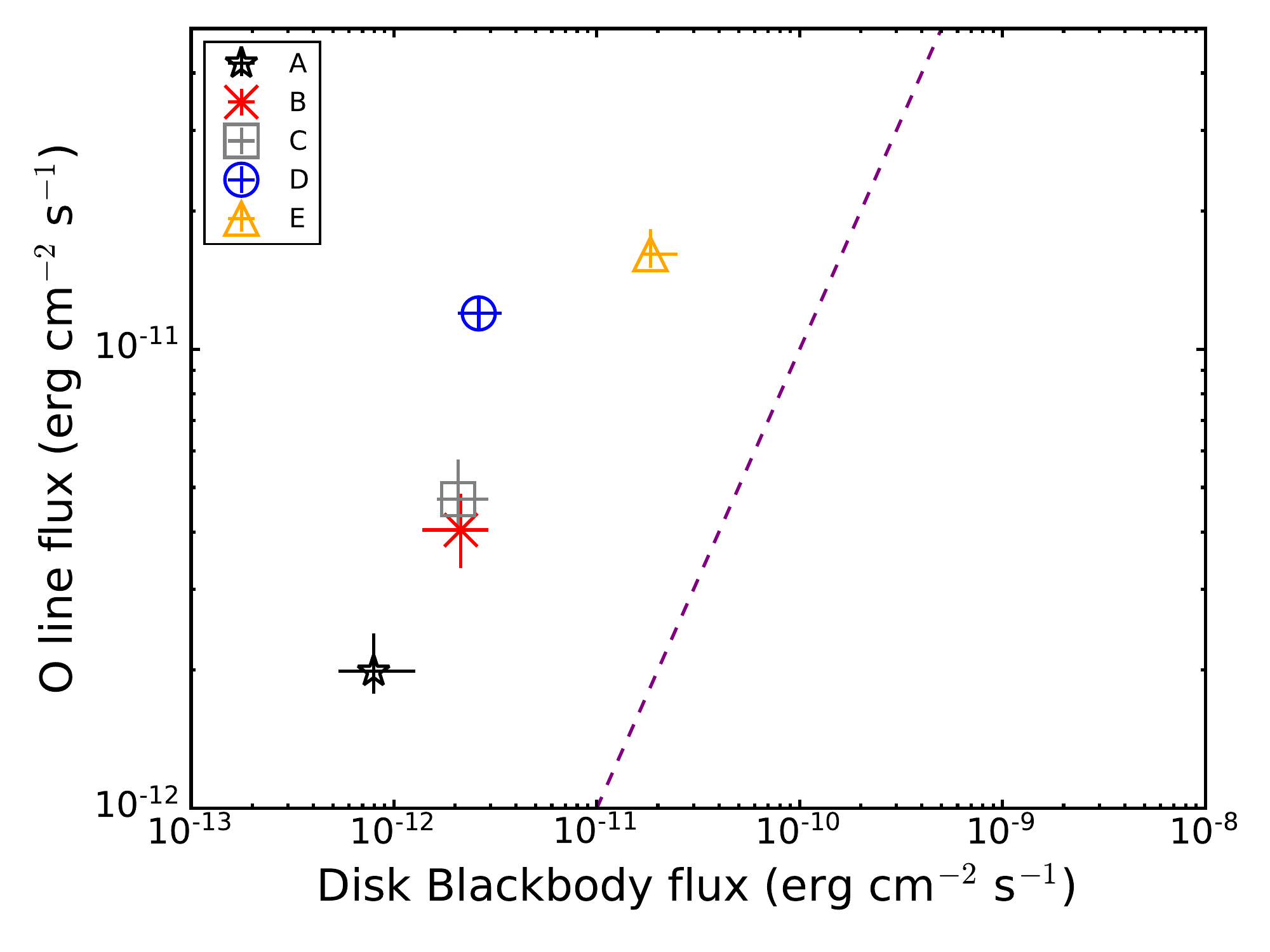}
\includegraphics[width=0.48\textwidth]{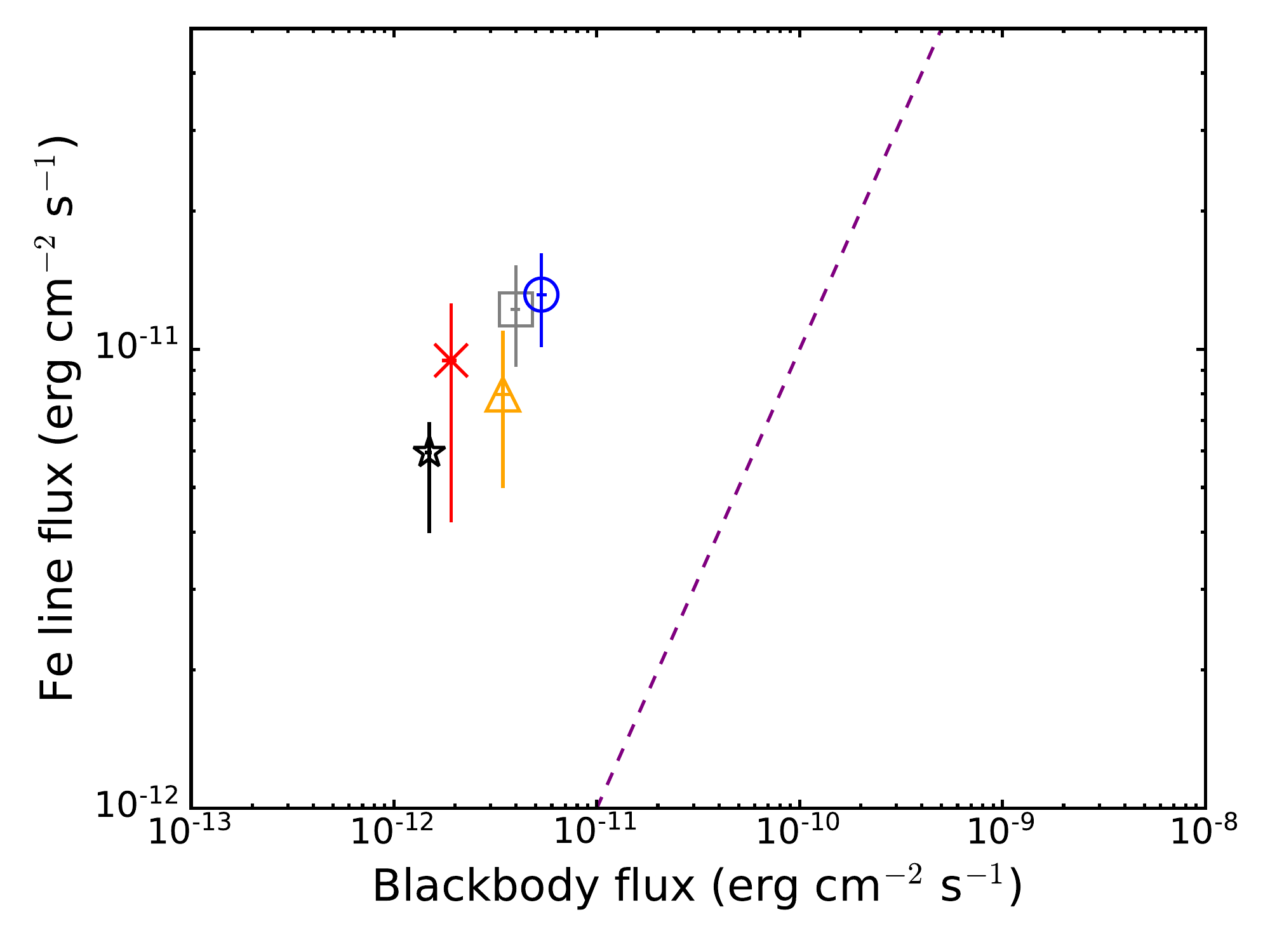}
\includegraphics[width=0.48\textwidth]{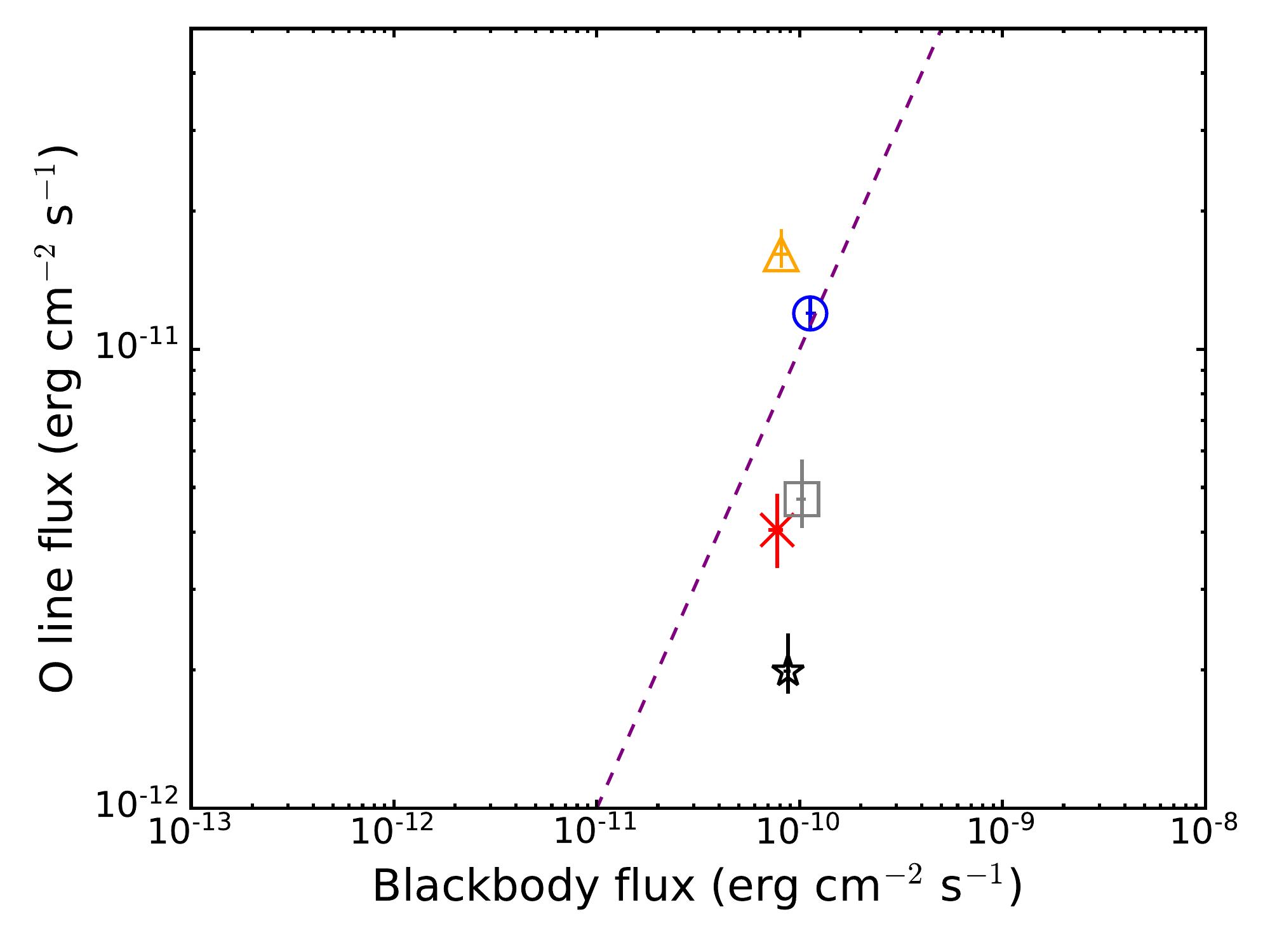}
\includegraphics[width=0.48\textwidth]{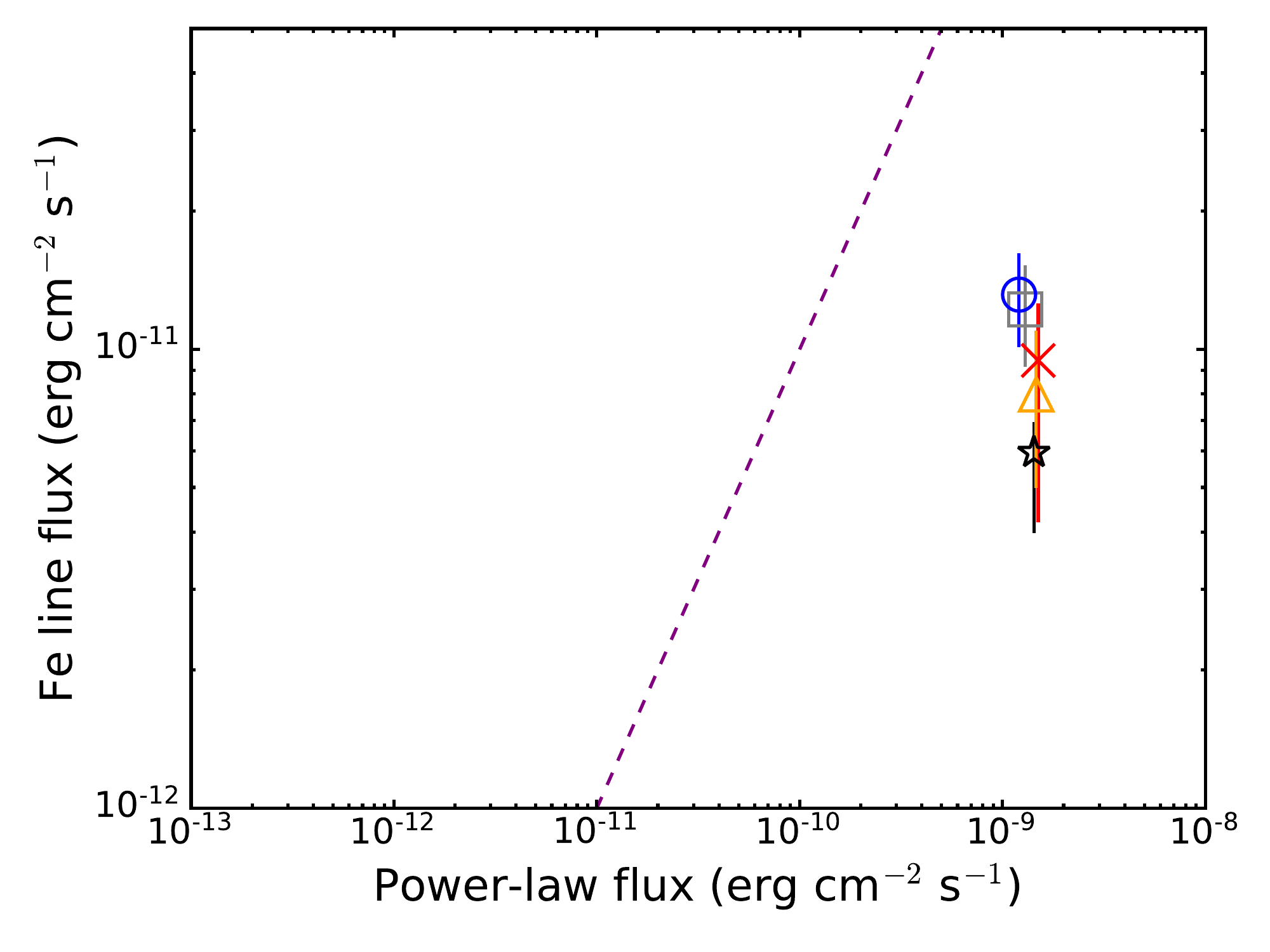}
\includegraphics[width=0.48\textwidth]{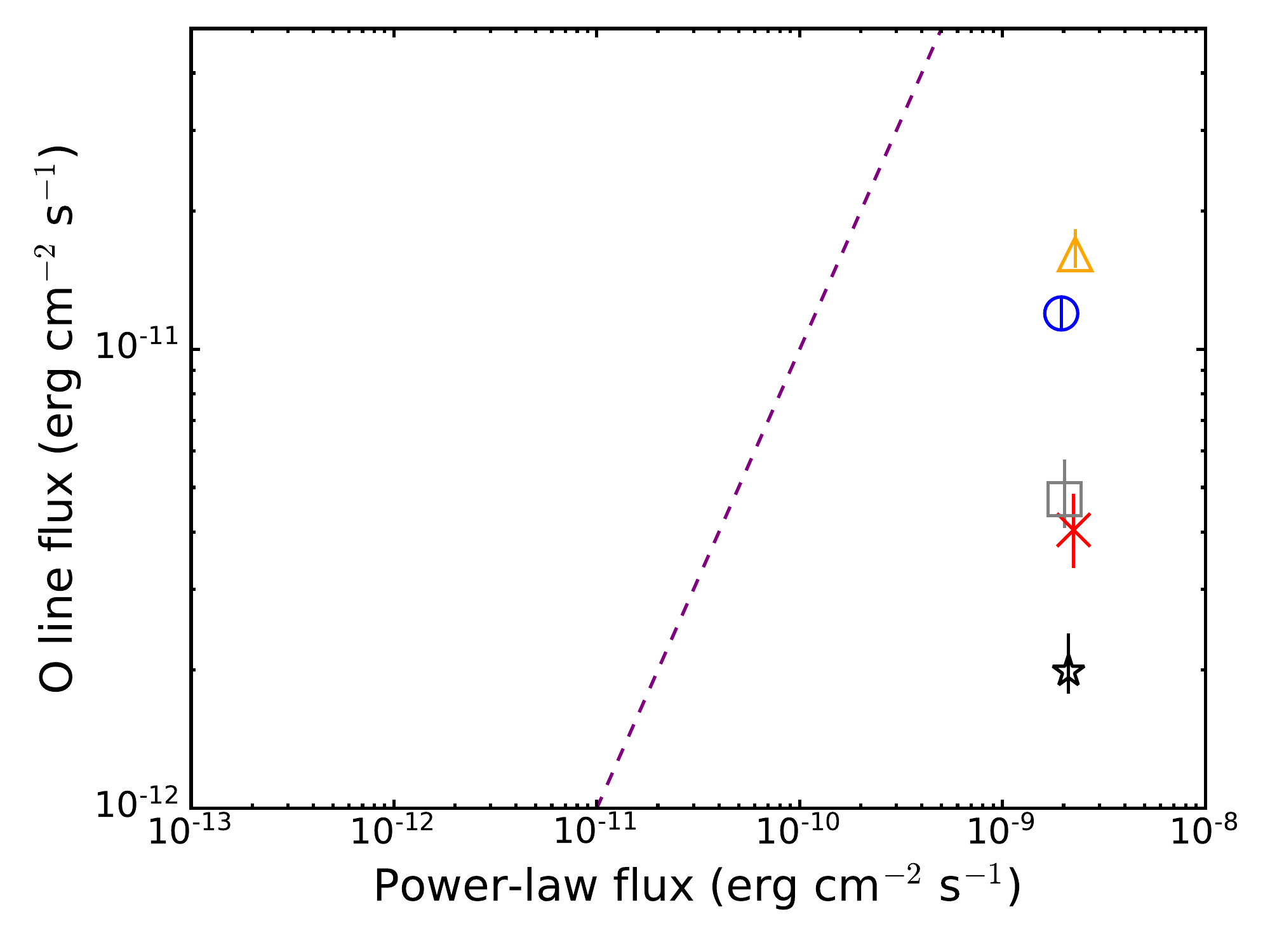}
\includegraphics[width=0.48\textwidth]{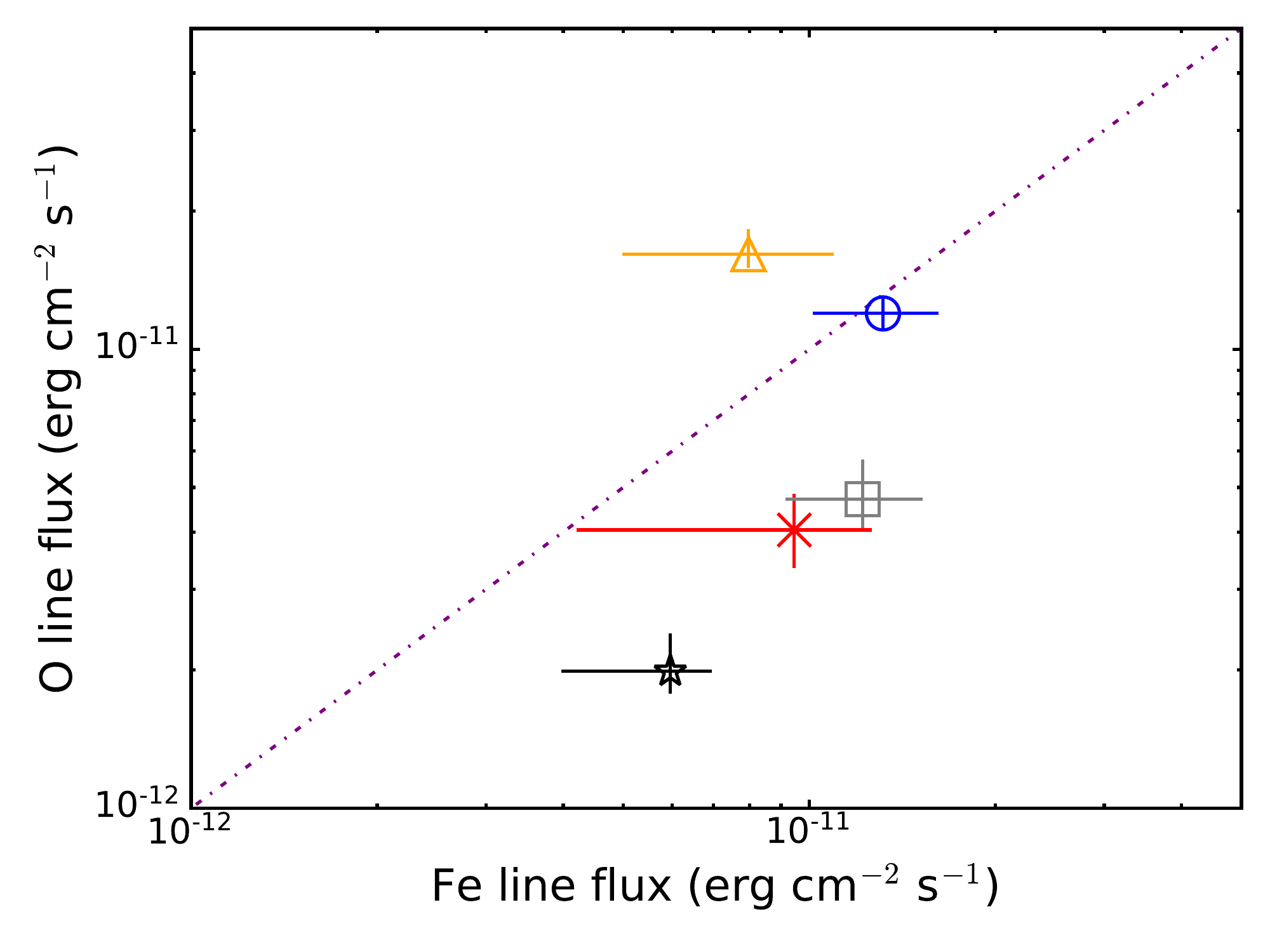}
\caption{Oxygen or iron line flux versus each continuum component for the five different time intervals shown in Figure \ref{fig:intervals}. The dashed purple line represents the trend expected if the O or Fe line flux is 10\% of the continuum component. 
The bottom right panel shows the O line flux versus the Fe line flux throughout the outburst; the dot-dashed line indicates a slope of one as reference. Both the O and Fe line fluxes increase with time, with the exception of Fe in interval E. }
\label{fig:flux}
\end{center}
\end{figure*}

We use {\sc tbabs} to account for the neutral absorption column along the line of sight, which is tied between intervals.
We concurrently fit the continuum for each spectrum using a variety of models that have been used previously for \source.
The parameter values and normalization of the continuum components are allowed to vary in order to account for changes in the spectral shape as the source intensity increases. 
Using an absorbed double thermal component model, {\sc tbabs}*({\sc diskbb}+{\sc bbody}), similar to \citet{ng10}, provides a reduced $\chi^{2}_{\nu}=3.51$ and fails to model the \integral\ component in interval E. If instead we use an absorbed power-law and blackbody model as per \citet{juett03}, this improves the fit to $\chi^{2}_{\nu}=2.11$. If instead we use a Comptonization component, {\sc nthcomp}, in place of the power-law similar to \citet{schultz03}, we obtain a comparable fit of $\chi^{2}_{\nu}=2.11$ from {\sc tbabs}*({\sc nthcomp}+{\sc bbody}), assuming that the seed photons originate from the accretion disk. Conversely, if we assume the seed photons originate from the blackbody component, {\sc tbabs}*({\sc nthcomp}+{\sc diskbb}), then the fit becomes worse ($\chi^{2}_{\nu}=3.66$).
An absorbed cut-off power-law and blackbody model from \citet{madej11} further improves the overall fit to $\chi^{2}_{\nu}=2.07$.

The model that is able to best describe the \nicer\ and \integral\ data is the hybrid model for NSs developed by \citet{lin07}. 
This is composed of a multi-temperature blackbody ({\sc diskbb}) for the accretion disk, a single-temperature blackbody ({\sc bbody}) for a boundary layer or emission from the surface of the NS, and a power-law component ({\sc powerlaw}) to account for emission from a coronal region.
We apply the same continuum description to each interval given that the hard color in Figure \ref{fig:intervals} shows very little change throughout the outburst. This continuum model provides a reduced $\chi^{2}_{\nu}$ of 1.78. 
Table \ref{tab:spec} provides the parameter values for continuum fitting (Model 1) for intervals A-E. Each of these components are statistically needed at $>27\sigma$ level of confidence via an F-test.

Switching out the single-temperature thermal component for Comptonization, {\sc nthcomp}, provided a significantly worse fit ($\chi^{2}_{\nu}=2.35$).
Moreover, the Comptonized component tended towards the shape of a blackbody with a high optical depth of $\tau\sim9$.  Alternatively, if we switch the power-law out for the Comptonization component, still assuming the seed photons originate from the single-temperature blackbody, we obtain a marginally better fit over Model 1 in Table 1 ($\chi^{2}_{\nu}=1.75$). However, the electron temperature tends to the hard limit of $kT_{e}=1000$~keV in all cases in order to properly describe the \integral\ data. Fixing $kT_{e}$ at lower values provides a worse $\chi^{2}$ and fails to fit the high-energy data from the \integral\ observation. For this reason, we proceed with our analysis using the continuum description of \citet{lin07}.

There is an excess in emission in the \nicer\ spectra between $0.6-0.7$ keV and $6.0-7.0$ keV that can be attributed to the O and Fe K emission lines that have previously been detected in \source. 
We do not detect any emission features in the  \swift/XRT spectrum due to the low signal-to-noise from the short exposure and modest count-rate. 
Figure \ref{fig:lines} shows the ratio of the \nicer\ data to the respective continuum model for intervals A-E.  
A clear evolution in the oxygen line profile with time can be seen, whereas any change in the Fe line profile between intervals is not significantly detected. 
This may be due to there being three times less \nicer\ collecting area in the Fe K band than in the O band.
In Figure \ref{fig:Otan}, we  show the change in the O line profile throughout the outburst by dividing the spectrum in each interval by the time-averaged spectrum. 
There is a  deficit of emission (with respect to the time-averaged profile) between $0.7-0.8$~keV that corresponds to the apparent O {\sc vii} edge in interval A that disappears as the source intensity increases. 
This is in agreement with the behavior shown in Figure \ref{fig:lines}. 
The disappearance of the O {\sc vii} edge may be due to a change in the ionization state of the material with time.

To assess the broadening of the features between $0.6-0.7$ keV and $6.0-7.0$ keV, we employ simple Gaussian components. The line widths are between $1.6\times10^{-2}\ \mathrm{keV}<\sigma_{O}<10.2\times10^{-2}$ keV and $0.74\ \mathrm{keV}<\sigma_{Fe}<1.89$ keV, respectively. The O line width agrees with the values previously reported in \citet{madej11}, while the Fe component values are typical of the broadening seen in other accreting NS LMXBs (e.g., \citealt{cackett12}).
In order to obtain physical constraints from these features, we add two {\sc diskline} \citep{fabian89} components to model the emission line broadening due to accretion around a compact object with dimensionless spin of $a=0$ (where $a=cJ/GM^{2}$).
The inclination ($i$) and emissivity index ($q$) are tied between all line components  for all intervals.
The outer disk radius is fixed at 1000 \rg.
The line energy, inner disk radius, and normalization for each line are allowed to vary between intervals in order to track any qualitative changes in the accretion disk.
This provides a significant improvement in the overall fit ($\Delta\chi^{2}=2402.95$ for 32 dof).
Values for each parameter are presented under Model 2 in Table~\ref{tab:spec}.
Figure \ref{fig:spectra} shows the broadband spectral model and residuals for interval E. This is representative of the overall model applied to each interval, but we only show one epoch for clarity.

The line energy for O is slightly lower than the laboratory value for O {\sc viii} Ly$\alpha$ ($0.654$ keV), but this could be due to a blending with O {\sc vii} emission since \nicer\ has an energy resolution of $\sim80$ eV at 1 keV and/or partially due to the current uncertainty in the gain ($<10$ eV). 
The Fe line component is consistent with a blend of Fe {\sc xxv} with lower ionization states in intervals A, B, and potentially E. Intervals C and D are inconsistent with He-like Fe {\sc xxv}. This change in the ionization state of Fe line may be due to the spectral softening of the power-law component with time. This can lead to less Fe-ionizing flux in the harder energy band. The O line component requires less energy to become ionized and therefore remains consistent with the H-like charge state throughout the outburst with some contribution from He-like O early on. 

The inner disk radius appears to move inward as the intensity increases from interval A$\rightarrow$E. 
Although the radii inferred from the O feature and Fe feature differ, they agree at the $3\sigma$ level of confidence, again indicating an evolution in \rin. 
Figure~\ref{fig:EE0} shows the O and Fe line profile in velocity space during interval C.
These profiles exhibit comparable broadening, suggesting that they do indeed arise from a similar location in the accretion disk. 
We only show a single interval for clarity, but the other intervals show analogous behavior.
The inclination is between $i=70^{\circ}-80^{\circ}$, which is consistent with the range of inclination values obtained in \citet{madej11} and \citet{madej14}. 
 We also tie \rin\ between each {\sc diskline} component (Model 3 in Table~\ref{tab:spec}). The fit is $3.9\sigma$ worse than when we allow the O and Fe components to vary in radius (Model 2), but the inclination and overall trend of the inner accretion disk moving towards the compact object remains. Rather than asserting that these lines arise from the same area of the disk, we conservatively proceed with the individual measurements, though this does not exclude the possibility that the lines originate from the same region.

We find a multiplicative constant between \nicer\ and \swift\ of $\mathrm{C}=0.93\pm0.03$ for each model in Table~\ref{tab:spec}. The constant between \nicer\ and \integral\ is smaller ($\mathrm{C}\sim0.30-0.50$), but this is likely due to detector-based flux offsets. The low value could also partially be due to spectral evolution between the two fitting bands.  However, alternative models, such as a broken or cutoff power-law, do not offer statistical improvements and yield some fairly extreme parameters. 
There are currently no direct measurements of the cross-calibration constant between \integral\ and \nicer.  Future investigations using simultaneous observations of sources with \nicer, \nustar, and \integral\ will likely be able to provide a better determination of the cross-calibration constant since there would be joint energy coverage between all three missions.

Since the oxygen line profile shows a clear evolution with time, we examine the O line flux versus continuum component flux in intervals A-E in order to discern what the line is responding to within the system, similar to what was done in \citet{lin10} for 4U~1705$-$44. 
Figure \ref{fig:flux} shows the bolometric oxygen line flux versus the unabsorbed $0.57-100$~keV flux for the multi-temperature disk blackbody, single-temperature blackbody, and power-law component. 
This energy band is chosen to encompass all ionizing radiation above the ionization threshold energy for O {\sc vii} that could possible contribute to the flux of the line.
The purple dashed line indicates the ratio of 10\% of the line flux to the respective continuum component. 
The disk blackbody component shows a positive trend with the 
oxygen line flux as the source increases in intensity.

A simple Spearman rank test, a measure of the statistical dependence between the line flux and continuum component, returns a correlation coefficient of 0.9 (with an error of 0.3 and $p$-value of 0.037). 
While the $p$-value is not very significant, as it is limited by the available points, 
the other components have a Spearman rank coefficient consistent with zero and high $p$-values. 
This implies that the change in the O line shape is correlated with changes in the accretion disk component,  perhaps due to the combination of (1) a change in the ionization state of the material as the accretion disk temperature increases (see Table \ref{tab:spec}), (2) more line emission as the disk moves inwards and a larger surface area is able to be irradiated by the external ionizing flux, and/or (3) a change in accretion geometry.

We also examine which continuum components the Fe line is responding to throughout the outburst. We use the unabsorbed flux in the $6.4-100$ keV band for each continuum component. 
The thermal disk component does not contribute to the flux in the Fe band, therefore we only examine the correlation with the single-temperature blackbody and power-law in Figure \ref{fig:flux}.
The bolometric flux estimates for the broadened Fe K component are more uncertain than for the O line because fewer counts are available.
There appears to be a smaller change in the Fe flux over the different intervals, but the Fe line is more correlated with the blackbody component than the changes in the power-law. 
Moreover, Figure \ref{fig:flux} shows a comparison of the change in the O line flux and Fe line flux throughout the outburst. 
There is an overall increase in bolometric line flux for each as the source intensity increases, with the exception of the Fe component in interval E. 
In any case, the O component serves as a better diagnostic tool for \nicer\ in UCXBs since there is a smaller signal-to-noise ratio in the Fe K band as the effective area declines rapidly, as well as the attenuation of Fe K emission by the overabundance of O in the accretion disk material \citep{koliopanos13}. 

The physical conditions of the line-emitting plasma can be characterized by the ionization parameter $\xi=L/n r^2$, where $L$ is the ionizing luminosity from $1-1000$~Ryd, $n$ is the number density of the plasma, and $r$ is the distance of the plasma from the ionizing radiation source taken to be $R_{in}$.  We can estimate a plausible change in $\xi$ at the beginning and end of the 10~day \nicer\ monitoring period studying the evolution of the O emission features. 
 Figure~13 of \citet{kallman01} shows the ionization balance of an optically-thin photoionized plasma at high density, as a function of $\xi$. Although we expect the accretion disk to be optically thick so we cannot use the exact values for $\xi$, this still allows us to qualitatively estimate how much $\xi$ should change in order to produce the evolution in spectral features that we observe. 
Given (1) the presence of an \ion{O}{7} edge while \ion{O}{8} is the dominant ion in interval~A and (2) that \ion{O}{7} is no longer present by interval E and the Fe K line is formally consistent with a blending of ionized \ion{Fe}{25} with neutral species \citep{kallman01, kallman04}, we determine that $\Delta (\log\xi)\sim0.3-1.2$ (cgs). 

From the limit on $\Delta(\log\xi)$, we can look into the change in the density of the material in the line emitting region using the following equation: 

\begin{equation}
\Delta (\log n)\simeq\log \left( \frac{F_{E}}{F_{A}} \frac{ \xi_{A}}{\xi_{E}} \frac{r_{A}^{2}}{r_{E}^{2}} \right)
\end{equation}

\noindent The ratio of the unabsorbed ionizing flux in interval E to interval A is $F_{E}/F_{A}=1.22$. Using the lower limit on inner disk radius in interval A (188~\rg) and upper limit in interval E (8.7~\rg), as well as the estimate for the change in ionization, we determine that the density increased by $1.6-2.5$ orders of magnitude over the course of the outburst. This change in density is large but could still fall within the range of density assumed within reflection models ($n\sim10^{15}-10^{19}$: \citealt{garcia16}).

 Moreover, we can use the equation for density of a standard thin disk in \citet{frank02} as a cross check for the range of values obtained from Equation~1: $\rho=3.1\times10^{-8} \alpha^{-7/10} \dot{M}_{16}^{11/20} m_{1}^{5/8} R_{10}^{-15/8} f^{11/5}$ g cm$^{-3}$, where $\alpha$ is the viscosity parameter, $\dot{M}_{16}$ is the mass accretion rate in units of $10^{16}$ g s$^{-1}$,  $m_{1}$  is the mass of the compact object in solar masses,  $R_{10}$ is the radius in units of $10^{10}$ cm, and $f=[1-R_{NS}/R_{in}]^{1/4}$.
$\Delta (\log \rho)$ and $\Delta (\log n)$ should change by the same amount since they are related by a constant.  Assuming that the viscosity parameter ($\alpha$) is constant, the change in density becomes: 

\begin{equation}
\Delta (\log \rho)\simeq\log \left[ \left(\frac{F_{E}}{F_{A}}\right)^{11/20} \left(\frac{ f_{E}}{f_{A}}\right)^{11/5} \left(\frac{r_{A}^{2}}{r_{E}^{2}}\right)^{-15/8} \right]
\end{equation}

\noindent For a NS with a 10 km radius and the same values for flux and inner disk radius used in Equation 1, the density in the line emitting region increases by $\sim2.5$ orders of magnitude as
the disk moves closer to the NS and the system becomes
more luminous. This is independent of the change in ionization and in agreement with our previous estimate of $1.6-2.5$. 
It is important to note that Equation~1 depends on the estimated change in $\xi$ and strongly on the change in inner disk radius ($\propto r^{2}$). Therefore, these estimates should be regarded with a degree of caution. For example, if we instead use the values reported in Table~1 for the Fe line component of 28~\rg\ in interval A and 10.2~\rg\ in interval E, we obtain a smaller change of \mbox{$\Delta (\log n)\sim-0.23-0.66$}.

Regardless, without fully self-consistent reflection modeling, we are unable to place further constraints on the change in ionization or disk density.
We tried fitting the data with an updated version of \xillver\ that has more grid points, but have identified limitations within the model, such as a steep dependence on small changes in $kT$, that resulted in unsatisfactory fits ( $\chi_{\nu}^{2}>80$). Therefore, we chose to not include them here. We are currently revising and updating the existing \xillver\ models and their application will be performed in a future investigation utilizing a sample of UCXBs.

\subsection{Timing}
From a $0.3-9.5$ keV light curve with 1/8192-s time resolution, we construct averaged power spectra for each interval according to usual methods \citep[see, e.g.,][]{vdK89, bult18}, normalizing the power to units of fractional rms with respect to the total source count-rate. The resulting power spectra are fit with a sum of Lorentzian profiles \citep{belloni02} and labeled according to the atoll naming conventions \citep{vdk06}. Note that the frequencies are also consistent with QPOs seen in stellar mass BHs and a different naming scheme could have been applied \citep{klein08}. However, since \source\ has exhibited a tentative Type-I X-ray burst, we opt for interpreting these features in the context of a NS.

\begin{table}[t]
    \newcommand{\mc}[1]{{#1}}
    \centering
	\caption{%
    	Power spectrum fit parameters
		\label{tab:fit}	
	}
    \begin{tabular}{l c c c c }
	\tableline
    ~     & \mc{Frequency} &\mc{Quality}& \mc{Fractional} & \multirow{3}{*}{$\chi^2$ / dof} \\
    ~     & ~              &\mc{factor} & \mc{amplitude}  &  \\
    ~     & \mc{(Hz)}      &\mc{~}      & \mc{(\% rms)}   &  \\
    \tableline
    \multicolumn2l{Interval A } \\
    \tableline 
    break2  & 0.16(0.03)   &  0.28(0.09)   &  6.2(0.8)   & \multirow{5}{*}{253 / 204} \\
    break   & 0.658(0.019) &  0.47(0.08)   &  12.4(0.6)  & \\
    hump    & 3.89(0.07)   &  0.59(0.05)   &  16.3(0.5)  & \\
    low     & 32.6(6.5)    &  0.0 ~(fixed)  &  12.1(0.6)  & \\
    hHz     & 143.8(3.9)   &  6.8(3.9)     &  4.9(1.0)   & \\
    \tableline 
    \multicolumn2l{Interval B } \\
    \tableline 
    break   & 0.87(0.05)  &  0.08(0.03)   &  15.5(0.4) & \multirow{4}{*}{224 / 208}\\
    hump    & 5.18(1.4)   &  0.58(0.08)   &  14.6(0.7) & \\
    low     & 42(12)      &  0.0 ~(fixed)  &  11.2(0.7) & \\
    hHz     & 245(12)     &  3.2(1.3)     &  7.1(0.9)  & \\
    \tableline 
    \multicolumn2l{Interval C } \\
    \tableline
    break   & 0.94(0.04)  &  0.10(0.02)   &  15.5(0.3)  & \multirow{4}{*}{264 / 185}\\
    hump    & 5.96(1.5)   &  0.56(0.07)   &  14.5(0.6)  & \\
    low     & 58.0 (4.0)   &  0.0 ~(fixed)  &  10.8(2.7)  & \\
    hHz     & 157.0 (23)   &  1.0(1.0)     &  8.0(4.0)   & \\
    \tableline 
    \multicolumn2l{Interval D } \\
    \tableline 
    break2  &  0.32(0.06) &  0.33(0.10)   &  5.9(1.1)  & \multirow{4}{*}{233 / 207}\\
    break   & 1.19(0.04)  &  0.44(0.10)   &  12.4(1.0) & \\
    hump    & 7.3(0.3)    &  0.29(0.06)   &  16.2(0.5) & \\
    hHz     & 367(12)     &  0.0 ~(fixed)  &  12.8(1.3) & \\
    \tableline 
    \multicolumn2l{Interval E } \\
    \tableline 
    break2  & 0.47(0.08)   &  0.27(0.08) &  6.5(0.9)    & \multirow{4}{*}{227 / 206}\\
    break   & 1.74(0.06)   &  0.54(0.11) &  11.3(0.9)   & \\
    hump    & 9.7(0.4)     &  0.33(0.08) &  14.7(0.6)   & \\
    hHz     & 223.0 (42)    &  0.4(0.3)   &  11.2(1.2)   & \\
	\tableline
	\end{tabular}
    \flushleft
    \tablecomments{%
     Best fit parameter values for the multi-Lorentzian 
     models describing the power spectra of \source. The values
     in parentheses indicate $1\sigma$ uncertainties. 
     }
\end{table}

\begin{figure*}[t]
  \centering
  \includegraphics[width=0.4\linewidth]{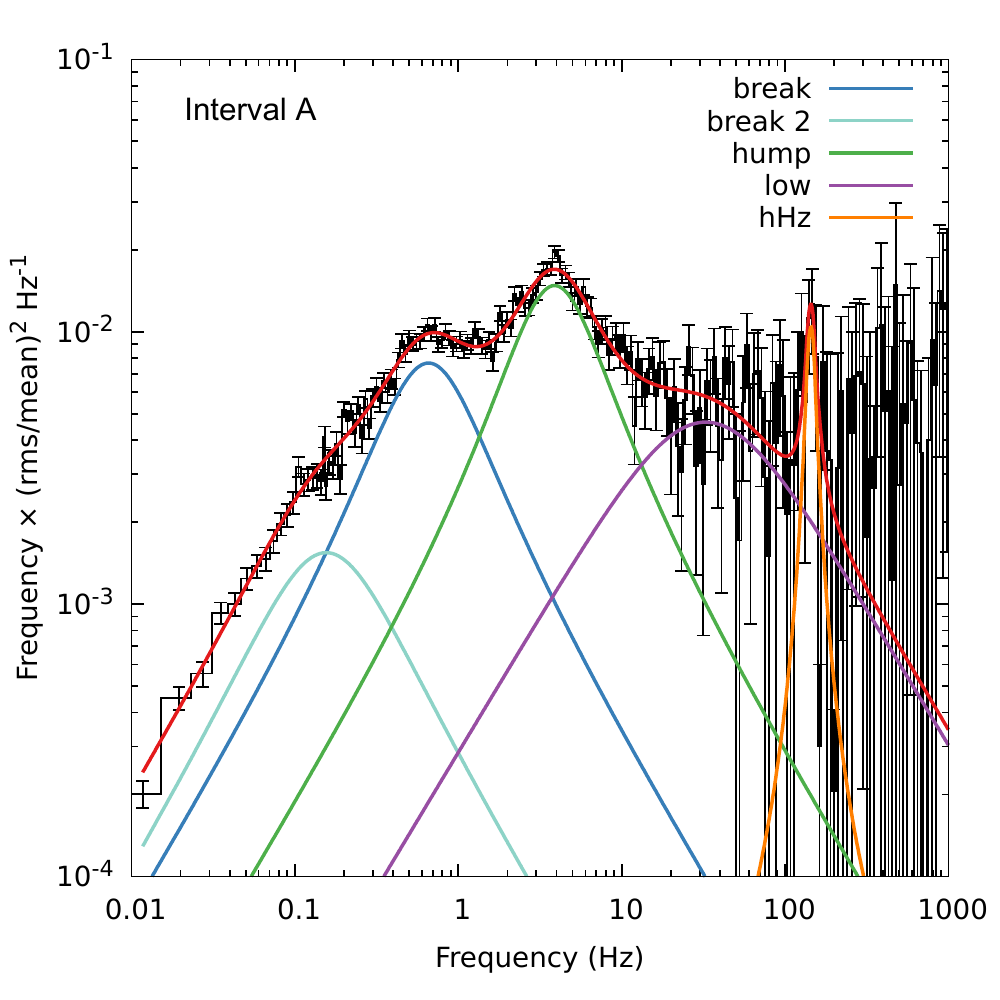}
  \includegraphics[width=0.4\linewidth]{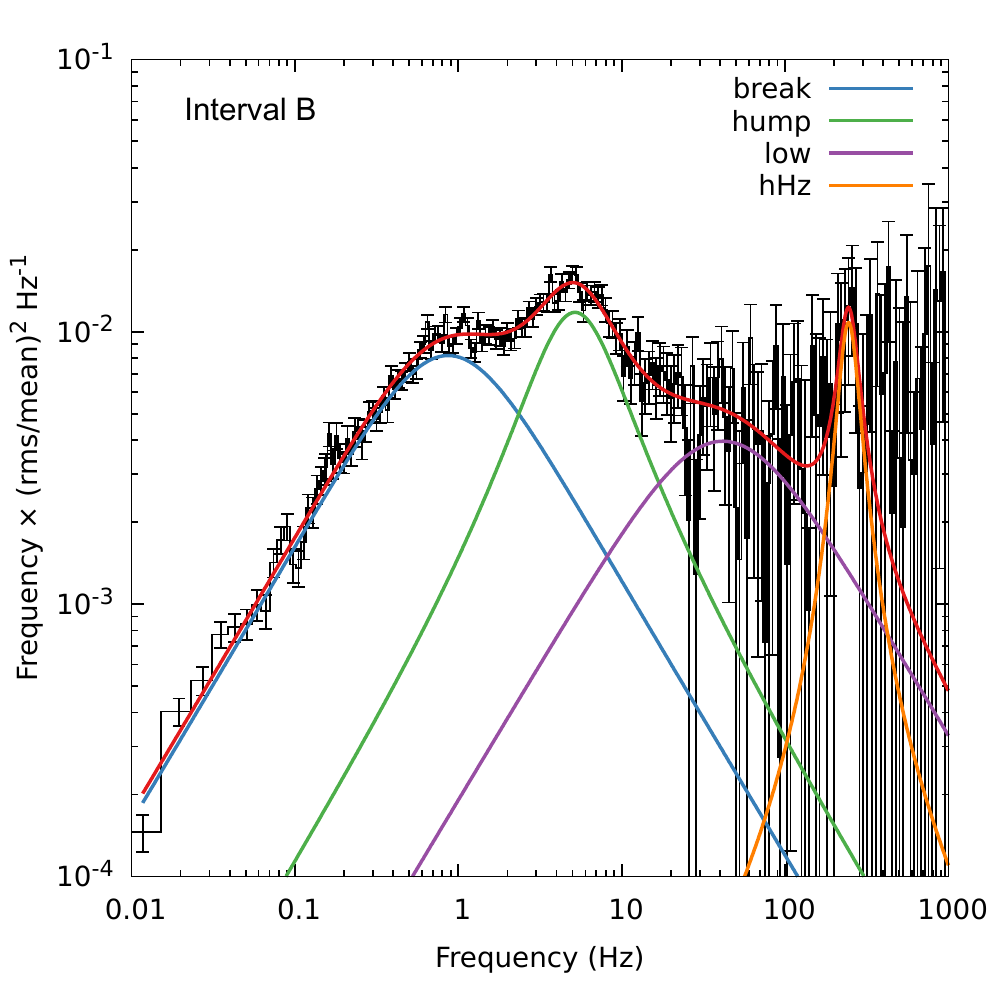}
  \includegraphics[width=0.4\linewidth]{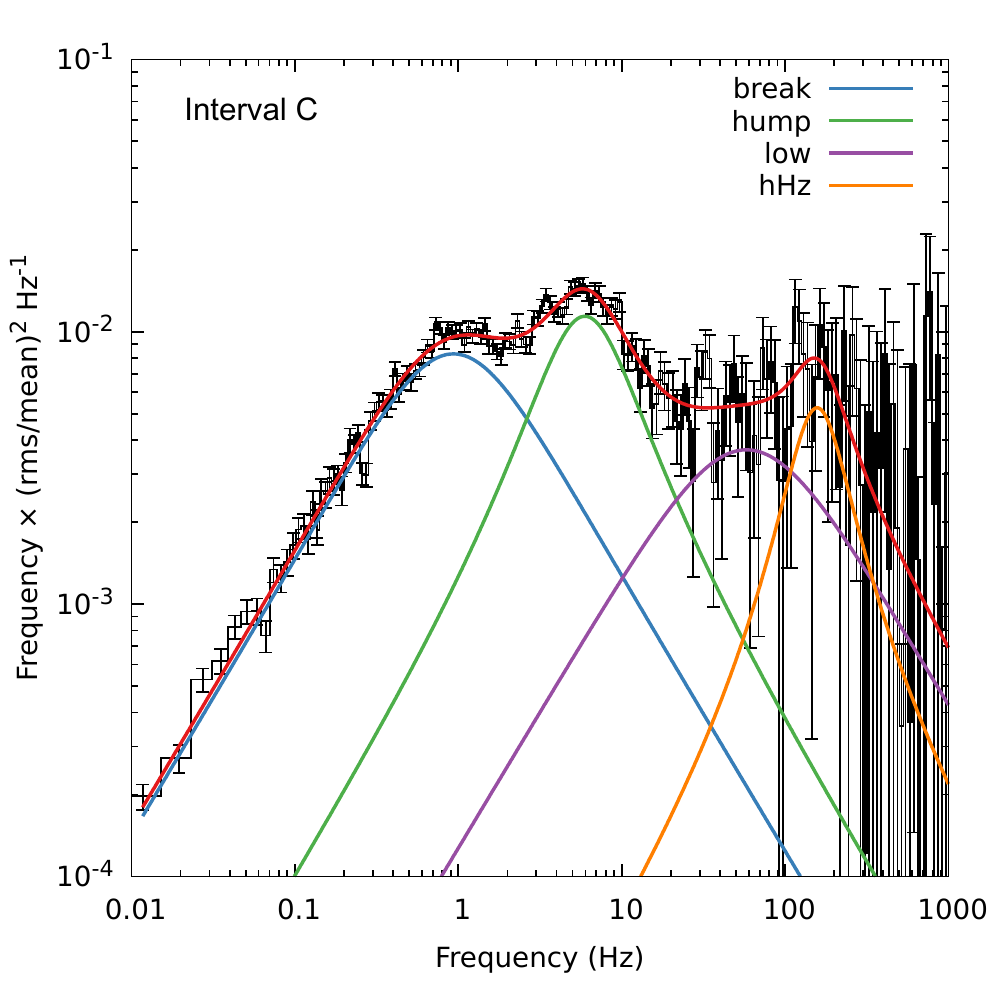}
  \includegraphics[width=0.4\linewidth]{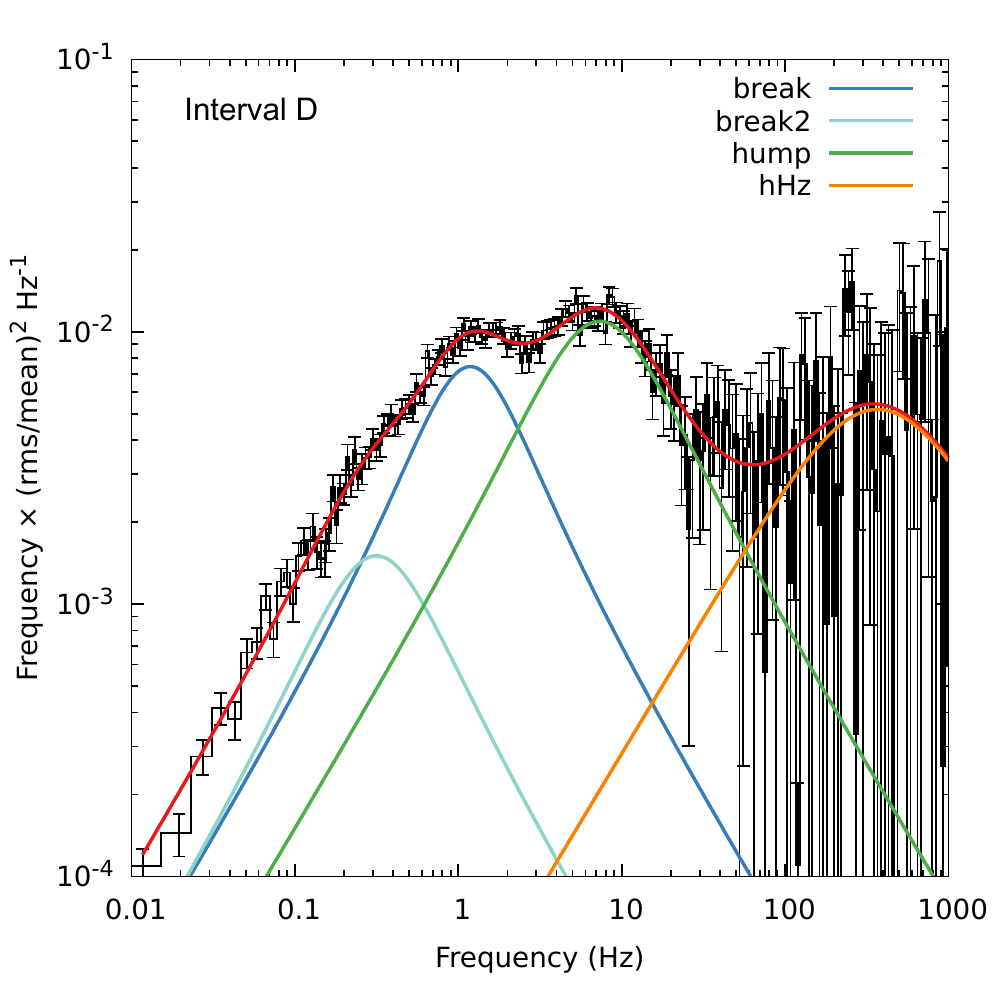}
  \includegraphics[width=0.4\linewidth]{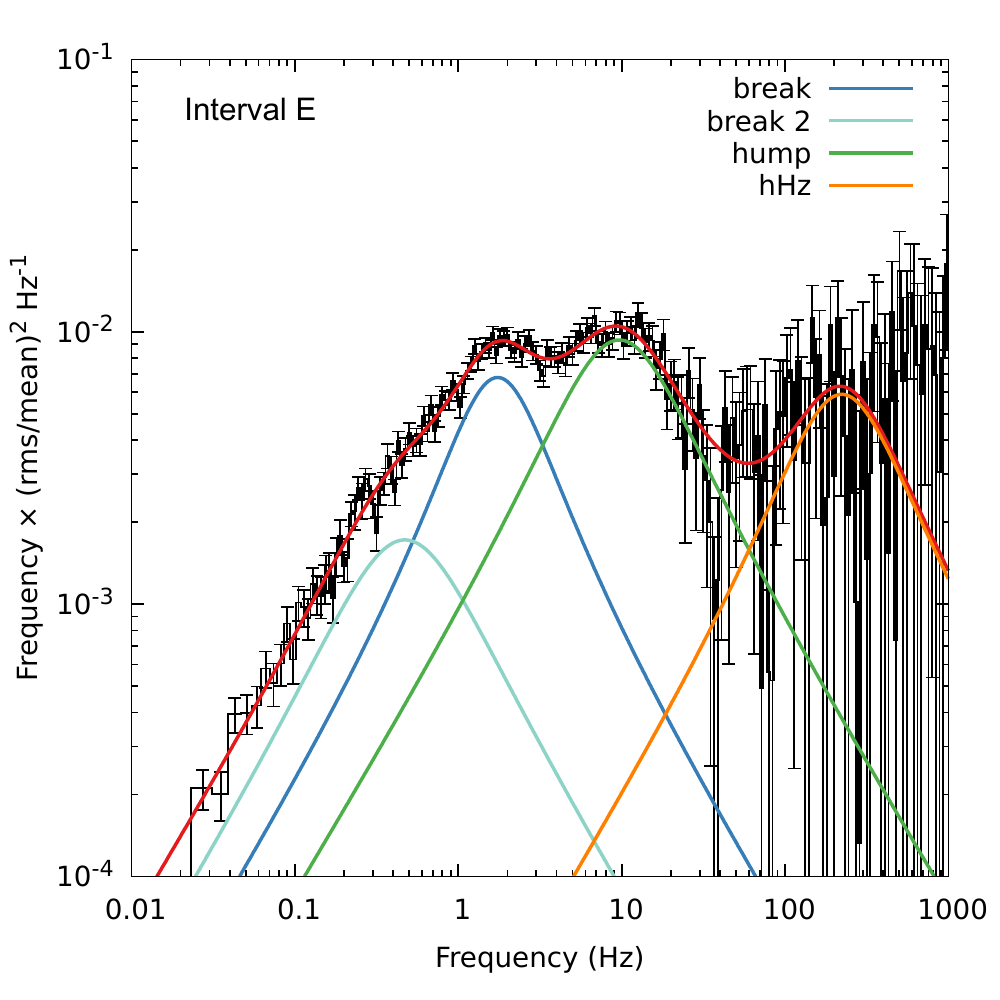}
  \includegraphics[width=0.4\linewidth]{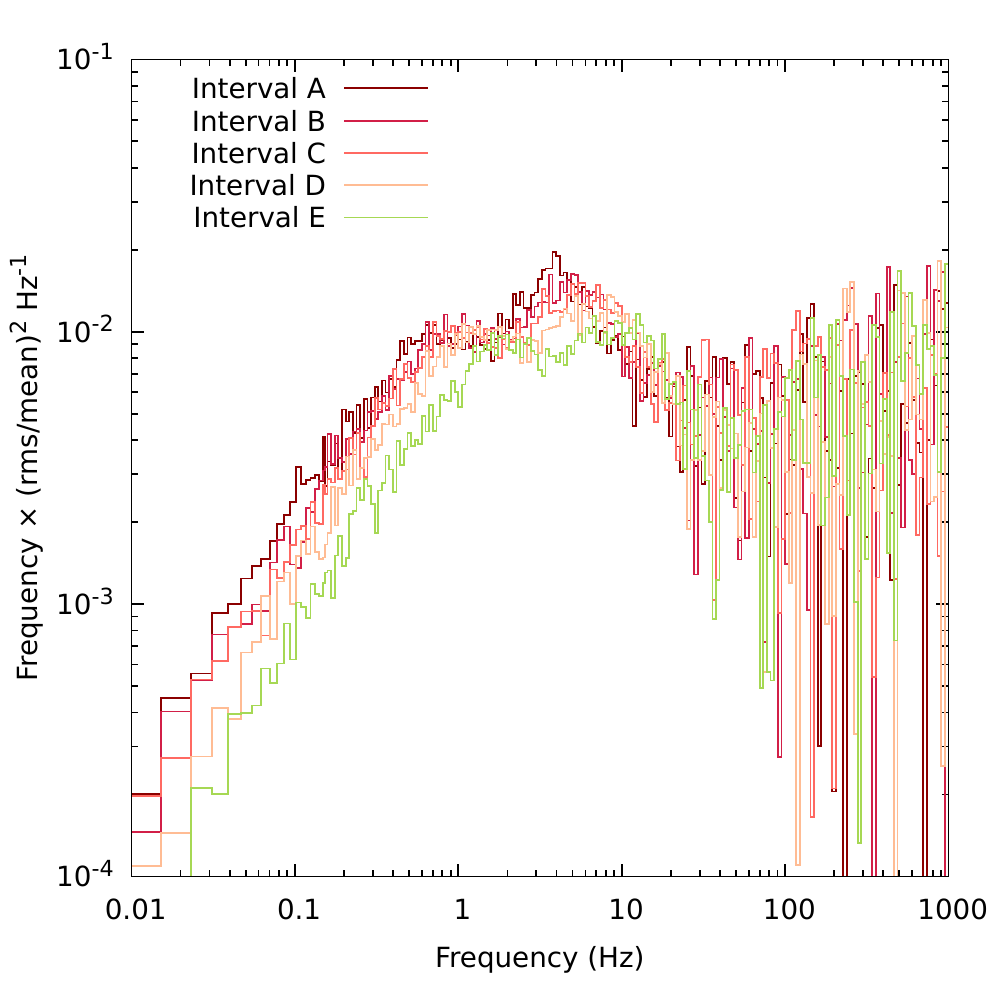}
  \caption{%
    Power density spectra of \source in intervals A-E with their best-fit model components, as well as a comparison of each interval. The power spectrum tends toward higher frequency and decreases in fractional rms amplitude as the source increases in intensity. This is in agreement with the behavior expected for an X-ray binary in the hard/intermediate state.
  }
  \label{fig:interval a-e}
\end{figure*}

\begin{figure*} 
\begin{center}
\includegraphics[width=0.98\textwidth]{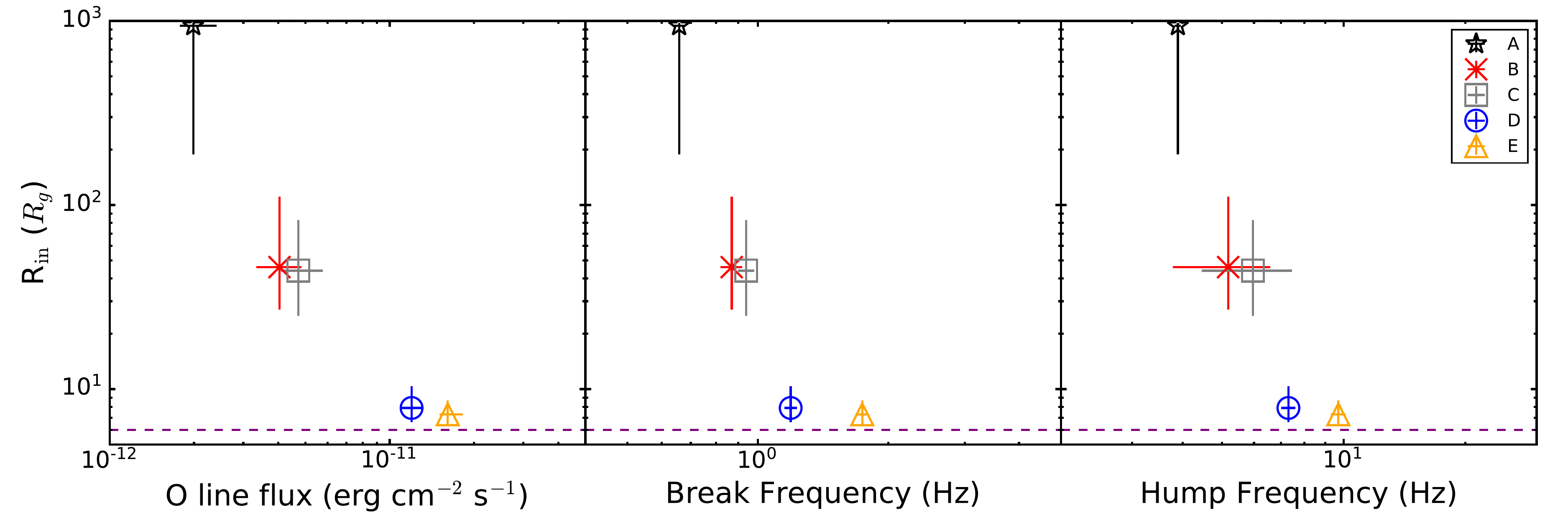}
\caption{The change in inner disk radius measured from the O line versus the O line flux, break frequency, and hump frequency throughout the outburst. Horizontal dashed purple line indicates 6 \rg\ or the innermost stable circular orbit for $a=0$. The inner accretion disk moves towards the compact object and the QPOs move to higher frequency as the source intensity increases (see  Figure 1).}
\label{fig:ORin}
\end{center}
\end{figure*}

We find that the power spectra of \source\ are well described by the sum of four or five Lorentzian profiles. 
The best-fit parameters are shown in Table \ref{tab:fit} and the individual power spectra and their best-fit model are shown in figure \ref{fig:interval a-e}. The narrow QPO seen at high frequencies in intervals A and B ($>100$ Hz, labeled `hHz' for hectohertz) is marginally significant at a detection level of $3.1$ and $3.0$ respectively. A visual inspection of interval D shows a 
similar narrow feature at a frequency of 250 Hz, however, with a signal-to-noise ratio of $\sim1.3$ this feature is not significantly detected. We find that as the source evolves to higher count-rate (interval A through E), the power spectrum components gradually move to higher frequencies, while decreasing in fractional rms amplitude. This trend is consistent with the expected behavior of a hard/intermediate state accreting X-ray binary.

Lastly, we searched for coherent pulsations in the data during the 10 day period of \nicer\ monitoring. 
This would provide further evidence that the compact object is a NS and classify the source as an X-ray pulsar. 
Both the individual and the total power spectra were searched, but no apparent coherent oscillations were found. 
It is possible that the Doppler modulation associated with the binary orbital motion suppresses the pulse signal to below our detection sensitivity \citep[see, e.g.,][]{strohmayer18}, however, a full acceleration search is beyond the scope of this work.

\section{Discussion}
We present evolution in the X-ray spectrum of \source\ during a period of enhanced accretion through \nicer\ monitoring over a $\sim10$ day period. 
We divided the data into five intervals that were labeled A-E. 
The soft color decreased with time while the hard color remained constant, suggesting an evolving low-energy thermal component within the system. 
Additionally, we obtained observations with \swift\ that were contemporaneous with interval A, as well as observations with \integral\ and \atca\ in interval E. 
The \integral\ observation played an important role in determining the appropriate continuum model for \source, since we were able to rule out models that were unable to describe the \nicer\ and \integral\ passbands simultaneously.

\begin{figure*} 
\begin{center}
\includegraphics[width=0.8\textwidth]{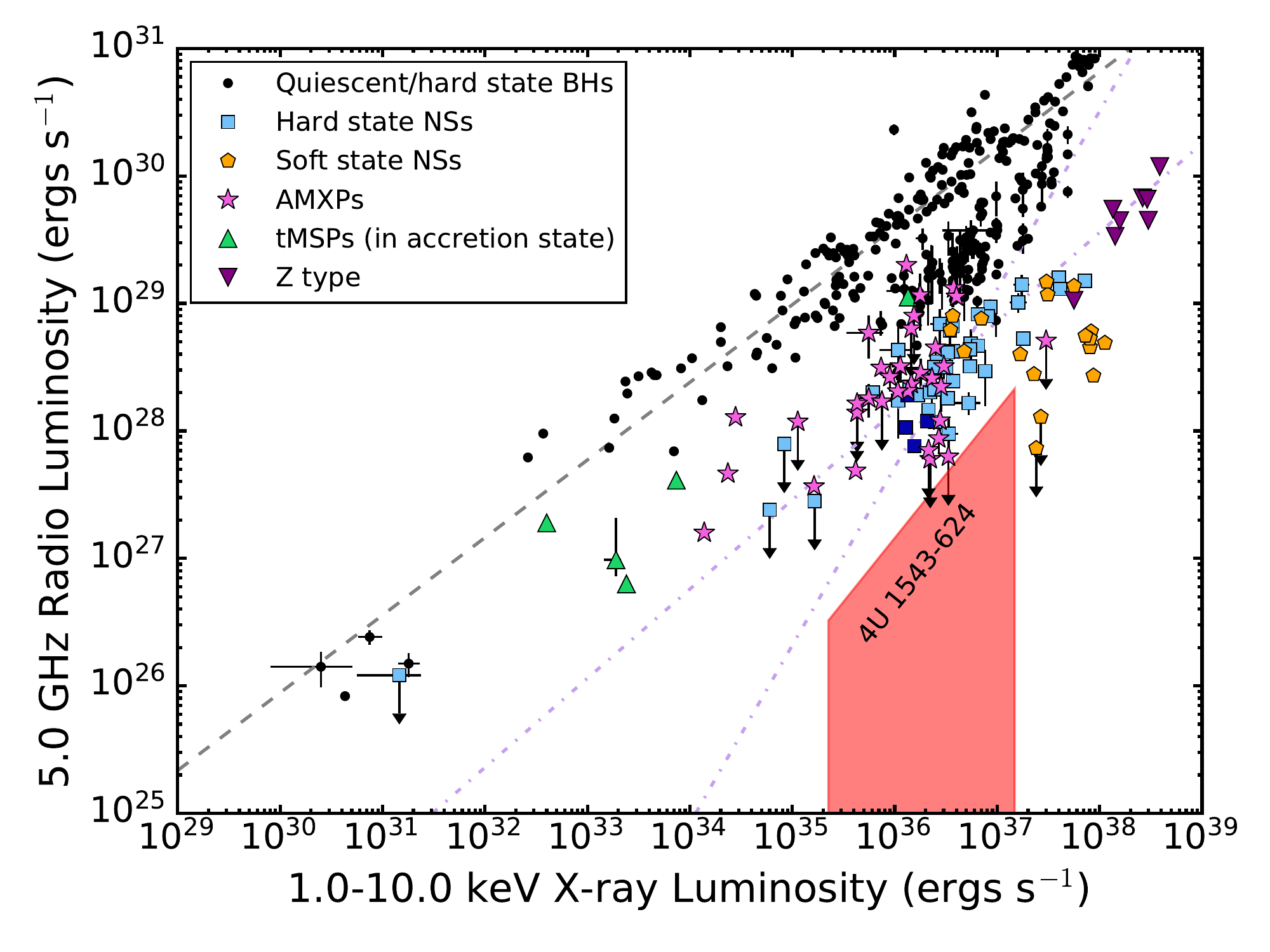}
\caption{Radio -- X-ray luminosity plane of accreting compact objects adapted from \citet{bahramian18}. The dashed grey line indicates the best-fit trends for BHs (${L}_{{r}}\propto{L}_{{x}}^{0.6}$; \citealt{gallo14}). The purple dot-dashed lines indicate the proposed correlations for NSs \citep{migliari06} for atoll and Z sources (${L}_{{r}}\propto{L}_{{x}}^{0.6}$) and for NSs in the hard state (${L}_{{r}}\propto{L}_{{x}}^{1.4}$). The black points are quiescent/hard state BHs from the available literature \citep{calvelo10, coriat11b, rodriguez11, paizis11, jonker12, corbel04, corbel13, brocksopp05, brocksopp13, gallo03, gallo14, russell15,  plotkin17, ribo17, dincer18}. The light and dark blue squares are NSs in hard states while orange pentagons are NSs in soft states (\citealt{rutledge98, migliari06, tudose09, migliari11, coriat11, tetarenko16, tudor17, gusinskaia17}); the dark blue is used to differentiate the UCXB 4U 0614+091 from the other hard state NSs. The magenta stars are accreting millisecond X-ray pulsars (AMXPs: \citealt{migliari05, migliari11, tudor16, tudor17, tetarenko17, vdE18a, strohmayer18a}). The green triangles are transitional millisecond pulsars (tMSPs) when they have switched from rotation powered to accretion powered mode \citep{hill11, papitto13, deller15, bogdanov18}. The purple upside-down triangles are Z sources \citep{migliari06}. The shaded red region indicates where \source\ resides based upon the X-ray flux in interval E, the 27 $\mu$Jy upper limit on the radio flux density, and the uncertainty on distance. The source lies two orders of magnitude below the trend for black holes.}
\label{fig:lrlx}
\end{center}
\end{figure*}

The \nicer\ spectra exhibit emission lines from O at $\sim0.64$~keV and Fe~K at $\sim6.4$~keV throughout the outburst.
The O line showed a clear evolution in the shape of its profile as the source intensity increased. 
The Fe line profile does not appear to change as dramatically, which is likely due to a combination of (1) there being three times less \nicer\ collecting area in the Fe band in comparison to the O region and (2) the line emission being dampened from the overabundance in C/O from the companion \citep{koliopanos13}. 
Both show a similar line profile when plotted in velocity space, indicating that they arise from a similar location within the disk.
With the advantage of the large collecting area of \nicer\ below 1.5 keV, we are able to track changes in the accretion disk using the O line component. 
The innermost accretion disk moves from hundreds of \rg\ initially to $<8.7$ \rg\ at peak intensity. 

The evolution in the power spectrum is consistent with an X-ray binary in the hard/intermediate state. 
As expected, the QPO frequencies move as the source intensity changes \citep{hasinger89}, aside from the hHz feature which remains consistent within a factor of~2 \citep{vanstraaten03}. 
The upper kHz QPO is often associated with the inner accretion flow, and although we are unable to detect any kHz QPOs, it is known that the frequency of kHz QPOs moves in conjunction with the lower QPO features \citep{vdk04}. 
Since the QPO features are moving towards higher frequency with time and the inner disk radius is decreasing, it stands to reason that the process controlling the QPO frequency is moving with the inner radius. 
In this regard, the mechanisms producing the QPOs are very likely dependent upon the inner disk radius. 
However, without measuring the upper kHz QPO frequency we are unable to test if the kHz QPO and reflection features are produced in the same region of the accretion disk.
Figure~\ref{fig:ORin} shows the inner disk radius  inferred from {\sc diskline} fitting versus the change in the O line flux, break frequency, and hump frequency with time.

The \chandra\ observation of \source\ analyzed in \citet{madej11} and \citet{madej14} occurred at a similar flux as the \nicer\ observations at peak intensity (i.e., $F_{\mathrm{0.5-10.0\ keV}}\sim1\times10^{-9}$ \fluxcgs).
The inner disk radius inferred from the O line in the final two intervals (D: \rin\ $=7.9_{-1.3}^{+2.5}$~\rg, E: \rin\ $=7.3_{-1.0}^{+1.4}$~\rg) agree well with the measurements from \citet{madej14} using \chandra/LETGS and the initial grid of \xillver\ (\rin\ $<7.4$~\rg). 
The equivalent width of the O lines in interval D ($\sim27$~eV) and E ($\sim33$~eV) also agree with the range reported in \citet{madej11}  of $27-35$~eV.
Moreover, the inclination of $i\sim70^{\circ}-80^{\circ}$ agrees with the upper limits reported in both studies. 
Our results are subject to change as our understanding of the \nicer\ instrument calibration improves, but we expect that this will only have a minor impact and the agreement with previous studies is reassuring.

Although often assumed to be a NS, the identity of the central object of \source\ is ultimately unknown. 
Constraining the mass function of the binary could conclusively confirm the presence of a BH or massive NS, but would require repeated spectroscopy of the faint counterpart --- currently impossible in a very short period system. 
The Type-I X-ray burst of \citet{serino18} would confirm that the compact accretor is a NS, but the resolution of the MAXI instrument (1.5$^{\circ}$) precludes a definite association with \source\ as there is another known X-ray source within the MAXI beam.
X-ray pulsations would also establish that \source\ contained a NS, but none were found in our inspection of the \nicer\ data. 
Finally, inspection of our simultaneous \atca\ radio and \nicer\ observations in interval E could constrain the nature and behavior of \source\ in the context of the known X-ray and radio properties of other LMXB systems.

The empirically-derived \lrlx\ fundamental plane displays the relationship between radio and X-ray luminosity for BHs and different classes of NS LMXBs, and aids in the understanding of the inflow/outflow efficiency of these systems \citep{gallo18}. 
In order to place \source\ on the \lrlx\ plane, we need to convert the radio flux density and X-ray flux into luminosities, which depends on the distance to the source.
\citet{wang04} originally estimated a distance of $\approx7$ kpc by calculating a systemic mass accretion rate using the orbital period, and then scaling the corresponding X-ray luminosity to that previously observed \citep{juett03}. 
The Type-I X-ray burst reported in \citet{serino18} places the source at a distance of $9.2\pm2.3$ kpc, assuming the empirical Eddington limit of $3.8\times10^{38}$ erg s$^{-1}$ \citep{kuulkers03}. \citet{bailerjones18} infer a limit on the distance to the optical companion of \source\ of $d=3.3^{+3.3}_{-1.9}$ kpc using {\it Gaia} DR2 parallax measurements with the assumption of an exponentially decreasing space density prior \citep{bailerjones15, astraa16}. Given the uncertainty surrounding the source location, we choose to encompass the entire range of distance estimates, 1.4 kpc $<d<$ 11.5 kpc,  in our luminosity calculations.

Assuming a flat radio spectrum, we convert our flux density upper limit to its corresponding 5.0 GHz luminosity with \lr ~$= 4{\pi}{d^{2}}{\nu}{S_{\nu}}$, where ${S_{\nu}}=27\ \mu$Jy, $\nu=5.0$ GHz, and $d$ is the distance to \source. We then convert the \nicer\ unabsorbed $1.0-10$ keV flux of $\sim9.34\times10^{-10}$ \fluxcgs\ into an X-ray luminosity.
By comparing the X-ray--radio luminosity to other known accreting LMXBs, we can determine if the properties of \source\ closely resemble those of another class of LMXBs.
Figure \ref{fig:lrlx} shows the region (in red) that \source\ traces out on the \lrlx\ plane based upon the current estimates of the distance to the source.
It is immediately clear that \source\ has a radio luminosity that is (a minimum of) two orders of magnitude below what is typically observed for BHs.
Although the \lrlx\ plane should be used with caution in making definitive claims about an LMXB's nature (e.g., IGR~J17591$-$2342: \citealt{russell18}, \citealt{ferrigno18}), we consider its position to be highly suggestive that the central object is some class of NS.

The utility of the \lrlx\ plane for separating NSs by class is somewhat more limited. 
We consider a Z type origin unlikely as \source\ is at least a magnitude fainter in both radio and X-ray luminosity than the faintest Z source, and displays none of the fast X-ray variability typical in such systems (\citealt{migliari06}).
The X-ray properties of \source\ alone point away from a soft state system, as its spectrum has been observed to be persistently hard.
The behavior of accreting tMSPs is more difficult to characterize as there are few confirmed tMSP systems. 
Presently, tMSPs are  among the most radio-bright NS systems and, if they obey their current observed \lrlx\ trend \citep{deller15}, are expected to be at least an order of magnitude brighter in the radio than our upper limit for \source.
Furthermore, they exhibit many other unique phenomena, e.g., optical pulsations, radio pulsations, rapid X-ray and radio flaring, gamma ray emission --- none of which have been observed in \source\ (\citealt{hill11}; \citealt{papitto13}; \citealt{bog15a}; \citealt{bog15b}; \citealt{zampieri19}). \

To further aid our interpretation, we consider 4U~0614+091, another UCXB that exhibits similar X-ray spectral properties to \source. 
Both show the presence of a prominent O {\sc viii} feature and a broad Fe K line (\citealt{madej14}; \citealt{ludlam19}). 
Unlike \source, 4U~0614+091 has been detected in the radio band. 
The source is located on the \lrlx\ plane (dark blue squares in Figure \ref{fig:lrlx}) near ${L}_{r,\ 5 \ \mathrm{GHz}}\sim10^{28}$\,erg\,s$^{-1}$ and ${L}_{x,\ 1-10\ \mathrm{keV}}\sim10^{36}$\,erg\,s$^{-1}$ \citep{migliari05}.
For the same X-ray luminosity, \source\ would be more than five times less radio luminous than 4U~0614+091. The upper limit on the magnetic field in 4U~0614+091 is $B\leq14.5\times10^{8}$ G at the poles \citep{ludlam19}. 
Using the same procedure, we can estimate a conservative upper limit on the $B$-field in \source\ for comparison.
The unabsorbed flux in the $0.5-50$ keV band is $2.01\times10^{-9}$ \fluxcgs\ in interval E and the upper limit on \rin\ is 8.7~\rg.
This provides an upper limit of $B\leq3.2\times10^{8}$~G at $d=11.5$~kpc or $B\leq0.4\times10^{8}$~G at $d=1.42$~kpc.

The difference in the magnetic field strength of \source\ and 4U~0614+091 could contribute to the discrepancy in radio luminosity between these two UCXBs, but how magnetic field strength affects radio emission in NS binaries is poorly understood. 
Recent studies of both AMXPs and non-pulsating NS LMXBs do not find obvious trends in their radio emission with respect to magnetic field strength or other systemic properties (\citealt{tudor17}; \citealt{tetarenko18}), although AMXPs overall have slightly higher radio luminosities than typical NS systems. 
It is suggested that an interplay between NS spin and magnetic field strength could have an important role in the efficiency and method of jet production \citep{vdE18}, but further investigation on this point is required.

Although the absence of X-ray pulsations in \source\ favors a typical hard state NS scenario, we cannot rule out that the source contains an AMXP whose pulsations are shielded by the accretion geometry \citep{lamb09} or suppressed below our detection threshold by the short-period orbital motion  \citep{strohmayer18}. 
Regardless, the radio and X-ray relationship of \source\ strongly favors a NS over a BH and adds to the evidence that NSs do not follow a single track in the \lrlx\ plane, limiting its use in distinguishing between different classes of NSs based on radio and X-ray observations only.

\acknowledgements{ This work was supported by NASA through the NICER mission and the Astrophysics Explorers Program, and made use of data and software provided by the High Energy Astrophysics Science Archive Research Center (HEASARC). The Australia Telescope Compact Array is part of the Australia Telescope National Facility which is funded by the Australian Government for operation as a National Facility managed by CSIRO. R.M.L.\ acknowledges partial funding from a NASA Earth and Space Science Fellowship. E.M.C.\ gratefully acknowledges support through CAREER award number AST-1351222. M.F. acknowledges ASI financial/programmatic support via ASI-INAF agreement n. 2013-025.R1 and ASI-INAF N. 2017-14-H.0. J.A.G.\ acknowledges support from Chandra Theory grant TM8-19003X and from the Alexander von Humboldt Foundation. J.C.A.\ M.-J.\ is the recipient of an Australian Research Council Future Fellowship (FT 140101082), funded by the Australian Government.} \\

\end{document}